\RequirePackage{ifpdf}
\ifpdf 
\documentclass[pdftex]{sigma}
\else
\documentclass{sigma}
\fi

\newtheorem{result}{Result}[section]

\numberwithin{conjecture}{section}
\numberwithin{example}{section}
\numberwithin{lemma}{section}
\numberwithin{theorem}{section}
\numberwithin{definition}{section}
\numberwithin{corollary}{section}
\numberwithin{remark}{section}
\numberwithin{proposition}{section}
\newtheorem{claim}{Claim}[section]
\newtheorem{conclusion}{Conclusion}[section]

\numberwithin{equation}{section}

\begin{document}

\renewcommand{\thefootnote}{$\star$}

\renewcommand{\PaperNumber}{071}

\FirstPageHeading

\ShortArticleName{Einstein Gravity, Lagrange--Finsler Geometry, and Nonsymmetric Metrics}

\ArticleName{Einstein Gravity, Lagrange--Finsler Geometry,\\ and Nonsymmetric Metrics\footnote{This paper is a
contribution to the Special Issue ``\'Elie Cartan and Dif\/ferential Geometry''. The
full collection is available at
\href{http://www.emis.de/journals/SIGMA/Cartan.html}{http://www.emis.de/journals/SIGMA/Cartan.html}}}

\Author{Sergiu I. VACARU~$^{\dag\ddag}$}

\AuthorNameForHeading{S.I. Vacaru}

\Address{$^\dag$~The Fields Institute for Research in Mathematical Science,\\
\hphantom{$^\dag$}~222 College Street, 2d Floor, Toronto,\ M5T 3J1, Canada}
\Address{$^\ddag$~Faculty of Mathematics, University ``Al. I. Cuza'' Ia\c si, 700506, Ia\c si, Romania}
\EmailD{\href{mailto:Sergiu.Vacaru@gmail.com}{Sergiu.Vacaru@gmail.com}}
\URLaddressD{\url{http://www.scribd.com/people/view/1455460-sergiu}}

\ArticleDates{Received June 24, 2008, in f\/inal form October
13, 2008; Published online October 23, 2008}

\Abstract{We formulate an approach to the geometry of Riemann--Cartan spaces
provided with nonholonomic distributions def\/ined by generic of\/f-diagonal
and nonsymmetric metrics inducing ef\/fective nonlinear and af\/f\/ine
connections. Such geometries can be modelled by moving nonholonomic frames
on (pseudo) Riemannian manifolds and describe various types of nonholonomic
Einstein, Eisenhart--Mof\/fat and Finsler--Lagrange spaces with connections
compatible to a general nonsymmetric metric structure. Elaborating a
metrization procedure for arbitrary distinguished connections, we def\/ine the
class of distinguished linear connections which are compatible with the
nonlinear connection and general nonsymmetric metric structures. The
nonsymmetric gravity theory is formulated in terms of metric compatible
connections. Finally, there are constructed such nonholonomic deformations
of geometric structures when the Einstein and/or Lagrange--Finsler
manifolds are transformed equivalently into spaces with generic local
aniso\-tropy induced by nonsymmetric metrics and generalized connections. We
speculate on possible applications of such geometric methods in Einstein and
generalized theories of gravity, analogous gravity and geometric mechanics.}


\Classification{53A99; 53B40; 53C21; 53C12; 53C44; 53Z05; 83C20; 83D05; 83C99}


\renewcommand{\thefootnote}{\arabic{footnote}}
\setcounter{footnote}{0}

\section{Introduction}

The possibility that there are gravitational and matter f\/ield interactions
described by nonsymmetric metrics has attracted attention beginning 1925,
when A.~Einstein~\cite{nseinst1,nseinst} proposed an~unif\/ied theory of
physical f\/ield (there were considered nonsymmetric metrics and complex
tensor f\/ields with Hermitian symmetry). Then, at the second step, L.P.~Eisenhart investigated the properties of generalized Riemannin spaces
enabled with nonsymmetric metrics~\cite{nseisenh1,nseisenh2}. He put the
problem to def\/ine the class of linear connections which are compatible with
the symmetric part of a nonsymmetric metric (the Eisenhart problem). Further
developments extended the problem to compatibility with general nonsymmetric
metrics.

The third step of developments on nonsymmetric theory of gravity can be
associated with J.~Mof\/fat works \cite{moff1,moff1a} and further
modif\/ications proposed and outlined in \cite{moffrev,nsgtjmp},
generalization to noncommutative quantum gravity \cite{moffncqg}, and
applications in modern cosmology~\cite{moff0505326,prok,mofft}.

The mentioned Eisenhart problem was solved in an important particular case
in~\cite{nsmat} and than approached in a form when the compatibility is with
nonsymmetric metrics on generalized Lagrange spaces (extending the
Riemann--Finsler geometry) in \cite{mhat}, see a review of results, on our
conventional forth direction in geometry and physics with nonsymmetric
metrics, in Chapter~8 of monograph~\cite{ma1987}.

A set of additional arguments for geometric and physical models with
nonsymmetric metrics follows from the theory of nonholonomic Ricci f\/lows %
\cite{vrf01,vrf02} (the f\/ifth direction of research with nonsymmetric
metrics). It was mentioned that nonholonomic distributions on manifolds
result in nonholonomic deformations of geometric objects (local frames,
metrics, connections, functional and ef\/fective thermodynamical ef\/fects, \dots)
which def\/ine naturally new classes of canonical metric compatible linear
connections with nonsymmetric Ricci tensor. Such nonsymmetric sources in the
evolution equations for the nonholonomically constrained Ricci f\/lows, in
general, give rise to nonsymmetric components of metrics. Following certain
geometric methods of generating \mbox{exact} solutions for Ricci f\/lows of
physically valuable metrics in gravity theories \cite{vrfsol1,vvisrf1,vvisrf2,vrf04,vrf05}, we constructed in explicit form \cite{avnsm01} new classes of solutions def\/ining evolutions of Taub NUT,
solitonic and pp-waves and Schwarzschild metrics, in general relativity,
into nonsymmetric metrics. It was concluded that nonholonomic spaces with
nonsymmetric metrics and generalized connections arise naturally in modern
geometry and physics if we try to describe a constrained dynamics and f\/low
evolution of physical processes and geometrical/physical objects; a
rigorous study of such theories presents a substantial interest.

The aim of this work is to elaborate a geometric approach to physical models
and spaces enabled with nonholonomic (equivalently, nonintegrable/constrained) distributions and adapted frames with associated nonlinear
connection structures and nonsymmetric metrics. We shall formulate a
solution of the Eisenhart problem on nonholonomic manifolds\footnote{For instance, on (pseudo) Riemannian manifolds enabled with nonholonomic
distributions.} when modelling of various types of generalizations, including
those with nonsymmetric metrics and supersymmetric/spinor/noncommutative
variables of the Riemann--Finsler and Lagrange--Hamilton geometries, is
possible. This way, we shall provide a synthesis of the methods and ideas
developed in directions two, three and four (mentioned above) in a general
nonsymmetric metric compatible form, for various classes of linear and
nonlinear connection, in strong relations to the f\/ifth direction following
the methods of geometry of nonholonomic manifolds. As general references on
nonholonomic manifolds enabled with nonlinear connection structure, on the
geometry of spaces with generic local anisotropy and applications to modern
physics and mechanics, we cite the works~\cite{vrfg,ma}.

It should be emphasized that compatibility between the metric and general
connection structures is very important not only for modelling physical
theories in a more simplif\/ied geometrical form but also crucial for further
both conceptual and technical developments of spinor ana\-ly\-sis and theory of
Dirac operators, on nonholonomic spaces with nonsymmetric metrics. A spinor
formulation for nonsymmetric metrics and related nonholonomic manifolds will
allow us to establish a connection with the geometry of nonholonomic
Clif\/ford bundles and algebroids~\cite{vfs,vhs,vclalg}, noncommutative and/or
nonholonomic geometry and gravity \cite{vggr,vncg} and geometric
quantization~\cite{vqgr3} developed following the methods of Lagrange and
Finsler geometry and applied to the Einstein gravity and generalizations.

The work is organized as follows.
In Section~\ref{sec2} we outline the algebraic properties of nonholonomic manifolds
of even dimension enabled with nonlinear connection and nonsymmetric metric
structures. The basic results from the dif\/ferential geometry of
distinguished connections compatible with nonsymmetric metrics are provided
in Section~\ref{sec3}. A metrization method for distinguished connections with
nonsymmetric metrics is formulated in Section~\ref{sec4}. Section~\ref{sec5} is devoted to the
main theorems def\/ining the set of distinguished connections being compatible
with general nonsymmetric metric structures. In Section~\ref{sec6} we formulate the
nonsymmetric gravity in terms of nonholonomic frames with associated
nonlinear connection structure and distinguished linear connections; we
emphasize the possibility to perform all constructions in metric compatible
form. Some models of nonsymmetric Lagrange and Finsler spaces are analyzed
in Section~\ref{snsfls}, where there are considered the conditions when such geometries
can be generated by nonholonomic deformations of Einstein spaces and result
in analogous modelling of gravitational ef\/fects by corresponding
nonholonomic distributions. Conclusions and discussion are presented in
Section~\ref{sec8}. Some important local formulas are outlined in Appendix.

\textbf{Remarks on notations and proofs of results.}
In this work, one follows the conventions from \cite{vrfg}. The Einstein's
convention on summing ``up'' and ``low'' indices will be applied if the
contrary will not be stated. We shall use ``boldface'' letters, $\mathbf{A},
\mathbf{B}_{\ \beta }^{\alpha },\dots$ for geometric objects and spaces
adapted to (provided with) a nonlinear connection structure. In general,
small Greek indices are considered as abstract ones, which may split into
horizontal~(h) and vertical~(v) indices, for instance $\alpha =(i,a),\beta
=(j,b),\dots $ where with respect to a coordinate basis they run values of type
$i,j,\ldots=1,2,\dots,n$ and $a,b,\ldots=n+1,n+2,\dots,n+m,$ for $n\geq 2$ and $m\geq
1. $ There are introduced also left ``up'' and ``low'' labels of geometric
objects: for instance, ${}^{t}\varphi $ is the matrix transposed to a matrix
$\varphi$, and ${}_{n}\mathcal{R}=\{{}_{n}R_{\ \beta \gamma \tau }^{\alpha
}\},$ and ${}_{n}\mathrm{Ric}({}_{n}D)=\{{}_{n}R_{\ \beta \gamma }\}$ are
respectively the Riemannian and Ricci tensors for the normal d-connection ${}_{n}\mathbf{\Gamma }_{\ \beta \gamma }^{\alpha }.$ We shall omit certain
labels and indices if that will not result in ambiguities.

We shall not present detailed proofs if they can be obtained by local
computations similar to those presented in~\cite{vrfg,ma,ma1987,moffrev,nsgtjmp}. The main dif\/ference is that in this
article we work on nonholonomic manifolds with locally f\/ibred structure
def\/ined by the nonlinear connection structure but not on tangent bundles or
on usual Riemann--Cartan spaces and their nonsymmetric metric
generalizations. With respect to certain classes of associated nonholonomic
frames, the algebraic structure of formulas is similar to the case of
integrable nonholonomic distributions. This allows us to simplify
substantially the proofs of theorems even a formal dubbing into the
so-called horizontal and vertical components exists.

\section{Nonsymmetric metrics on nonholonomic manifolds}\label{sec2}

In this section we outline the algebraic properties of nonholonomic
manifolds of even dimension enabled with nonlinear connection and
nonsymmetric metric structures. We redef\/ine on such spaces with local f\/ibred
structure the constructions from~\cite{nsmat,mhat,ma1987}.

Let us consider a smooth manifold $\mathbf{V}^{n+n}$ of even dimension $n+n$
(in a particular case, for a~tangent bundle, $\mathbf{V}^{n+n}=TM,$ where $M$
is a smooth manifold of dimension~$n$). We denote local coordinates in the
form $u=(x,y),$ or$\ u^{\alpha }=(x^{i},y^{a}),$ where indices $%
i,j,k,\ldots =1,2,\dots,n$ and $a,b,c,\ldots=n+1,n+2,\dots,n+n$ (on $TM,$ we can use the
same indices for both base and f\/iber indices).

\begin{definition}
A nonlinear connection (N-connection) structure $\mathbf{N}$ is def\/ined by
a nonholonomic (nonintegrable) distribution (a Whitney sum)%
\begin{equation}
T\mathbf{V}^{n+n}=h\mathbf{V}^{n+n}\oplus v\mathbf{V}^{n+n}  \label{whitney}
\end{equation}%
into conventional horizontal~(h) and vertical~(v) subspaces.
\end{definition}

In local form, a N-connection is given by its coef\/f\/icients $N_{i}^{a}(u),$
for%
\begin{equation}
\mathbf{N}=N_{i}^{a}(u)dx^{i}\otimes \frac{\partial }{\partial y^{a}},
\label{coeffnc}
\end{equation}
stating on $\mathbf{V}^{n+n}$ a preferred frame (vielbein) structure
\begin{equation}
\mathbf{e}_{\nu }=\left( \mathbf{e}_{i}=\frac{\partial }{\partial x^{i}}%
-N_{i}^{a}(u)\frac{\partial }{\partial y^{a}},e_{a}=\frac{\partial }{%
\partial y^{a}}\right) ,  \label{dder}
\end{equation}%
and a dual frame (coframe) structure
\begin{equation}
\mathbf{e}^{\mu }=\left( e^{i}=dx^{i},\mathbf{e}%
^{a}=dy^{a}+N_{i}^{a}(u)dx^{i}\right) .  \label{ddif}
\end{equation}%
The vielbeins (\ref{ddif}) satisfy the nonholonomy relations
\begin{equation*}
\lbrack \mathbf{e}_{\alpha },\mathbf{e}_{\beta }]=\mathbf{e}_{\alpha }%
\mathbf{e}_{\beta }-\mathbf{e}_{\beta }\mathbf{e}_{\alpha }=w_{\alpha \beta
}^{\gamma }\mathbf{e}_{\gamma }  
\end{equation*}%
with (antisymmetric) nontrivial anholonomy coef\/f\/icients $w_{ia}^{b}=\partial
_{a}N_{i}^{b}$ and $w_{ji}^{a}=\Omega _{ij}^{a},$ where
\begin{equation}
\Omega _{ij}^{a}=\mathbf{e}_{j}\left( N_{i}^{a}\right) -\mathbf{e}_{i}\left(
N_{j}^{a}\right)  \label{ncurv}
\end{equation}%
are the coef\/f\/icients of N-connection curvature.  The particular holonomic/integrable case is selected by the integrability conditions $w_{\alpha \beta
}^{\gamma }=0.$\footnote{we use boldface symbols for spaces (and geometric objects on such spaces)
enabled with N-connection structure.}

\begin{definition}
A N-anholonomic manifold is a manifold enabled with N-connection structure~(\ref{whitney}).
\end{definition}

The nonholonomic properties of a N-anholonomic manifold are completely
def\/ined by the the N-adapted bases (\ref{dder}) and (\ref{ddif}). For
instance, we can always transform an arbitrary mani\-fold into a
N-anholonomic one by enabling it with a set of coef\/f\/icients $N_{i}^{a}$ for
a corresponding N-connection structure. Equivalently, we can say that such
coef\/f\/icients def\/ine  a class of (li\-near\-ly depending on $N_{i}^{a}$) linear
frames. A geometric/physical motivation for such constructions can be
provided if, for instance, the N-coef\/f\/icients are  induced by a (regular)
Lagrange structure (in geometric mechanics, see~\cite{ma,ma1987}), or
by certain of\/f-diagonal coef\/f\/icients of symmetric and/or nonsymmetric
metrics, see examples and details in~\cite{vrfg,vncg,vrf02,avnsm01}. A
manifold is not N-anholonomic if it is enabled with a trivial N-connection
structure when the corresponding nonholonomic coef\/f\/icients vanish.

One says that a geometric object is N-adapted (equivalently,
distinguished), i.e.\ a d-object, if it can be def\/ined by components adapted
to the splitting (\ref{whitney}) (one uses terms d-vector, d-form,
d-tensor)\footnote{As general references on the geometry of nonholonomic manifolds and
applications to modern physics, we cite \cite{vrfg,vncg} and, for former
constructions on tangent and vector bundles, \cite{ma,ma1987}. Readers may
consult those works and provided there references and appendices for details
and applications of the formalism of so-called d-tensors and d-objects.
We note that in this work similar constructions are generalized to the case
when manifolds (in general, nonholonomic ones) are enabled with nonsymmetric
metric structures and related linear connection structures.}. For instance,
a d-vector is written in the form $\mathbf{X}=X^{\alpha }\mathbf{e}_{\alpha
}=X^{i}\mathbf{e}_{i}+X^{a}e_{a}$ and a one d-form $\widetilde{\mathbf{X}}$
(dual to $\mathbf{X}$) is $\widetilde{\mathbf{X}}=X_{\alpha }\mathbf{e}%
^{\alpha }=X_{i}e^{i}+X_{a}e^{a}.$ On N-anholonomic manifold it is
convenient to work with d-objects because in this case all geometric and
physical objects are derived in a form adapted to the N-connection
structure (i.e.\ to the corresponding class of imposed nonholonomic
constraints). Of course, any d-tensor can be transformed into a general
tensor but in such a case we ``do not take care'' about existing N-connection
structure.

There is an almost complex structure $\mathbb{F}$ associated to a prescribed
N-connection structure $\mathbf{N}$, which is def\/ined by operators
\begin{gather}
\mathbb{F}\left(\mathbf{e}_{i} =\frac{\partial }{\partial x^{i}}-N_{i}^{a}\frac{%
\partial }{\partial y^{a}}\right)=-e_{i}=-\frac{\partial }{\partial y^{i}},
 \nonumber\\
\mathbb{F}\left(e_{i} =\frac{\partial }{\partial y^{i}}\right)=\mathbf{e}_{i}=\frac{
\partial }{\partial x^{i}}-N_{i}^{a}\frac{\partial }{\partial y^{a}}  \label{acs}
\end{gather}%
and has the property $\mathbb{F}^{2}=-I$, where $I$ is the identity matrix.

In this work, we study the geometric properties of spaces $(\check{g}_{ij},%
\mathbf{V}^{n+n},\mathbf{N}),$ where the h-subspace is enabled with a
nonsymmetric tensor f\/ield (metric) $\check{g}_{ij}=g_{ij}+a_{ij}$, where the
symmetric part $g_{ij}=g_{ji}$ is nondegenerated and $a_{ij}=-a_{ji}$.

\begin{definition}
A d-metric $\check{g}_{ij}(x,y)$ is of index $k$ if there are satisf\/ied the
properties:
\begin{gather*}
1)~\det |g_{ij}|\neq 0 \qquad \mbox{and}\\
2)~{\rm rank}\,|a_{ij}|=n-k=2p, \qquad \mbox{for}\quad 0\leq k\leq n.
\end{gather*}
\end{definition}

We denote by $g^{ij}$ the reciprocal (inverse) to $g_{ij}$ d-tensor f\/ield.
The matrix $\check{g}_{ij}$ is not invertible unless for $k=0.$

\begin{definition}
A N-anholonomic (we shall use also the term nonholonom\-ic) Eisenhart space
of index $k$ is a nonholonomic manifold $(\check{g}_{ij},\mathbf{V}^{n+n},\mathbf{N})$ provided with d-metric $\check{g}_{ij}=g_{ij}+a_{ij}$ of index~$k$.
\end{definition}

For $k>0$ and a positive def\/inite $g_{ij}(x,y),$ on each domain of local
chart there exists $k$ d-vector f\/ields $\xi _{i^{\prime }}^{i},$ where $%
i=1,2,\dots,n$ and $i^{\prime }=1,\dots,k$ with the properties%
\begin{equation*}
a_{ij}\xi _{j^{\prime }}^{j}=0\qquad \mbox{and}\qquad g_{ij}\xi _{i^{\prime }}^{i}\xi
_{j^{\prime }}^{j}=\delta _{i^{\prime }j^{\prime }}.
\end{equation*}%
If $g_{ij}$ is not positive def\/inite, we shall assume the existence of $k$
linearly independent d-vector f\/ields with such properties. For arbitrary
signatures of $g_{ij},$ we can chose any $k$ independent and orthonormalized
vectors def\/ined as a linear combination of a linear basis of $n$ vectors.

We note that we can def\/ine completely the metric properties on $\mathbf{V}%
^{n+n}$ if we state additionally that this space is provided with the metric
structure
\begin{gather}
\check{\mathbf{g}} =\mathbf{g}+\mathbf{a}=\check{\mathbf{g}}_{\alpha \beta
}\mathbf{e}^{\alpha }\otimes \mathbf{e}^{\beta }=\check{g}_{ij}e^{i}\otimes
e^{j}+\check{g}_{ab}\mathbf{e}^{a}\otimes \mathbf{e}^{b},\notag   \\
\mathbf{g}=\mathbf{g}_{\alpha \beta }\mathbf{e}^{\alpha }\otimes \mathbf{e}%
^{\beta }=g_{ij}e^{i}\otimes e^{j}+g_{ab}\mathbf{e}^{a}\otimes \mathbf{e}%
^{b},  \notag \\
\mathbf{a} =a_{ij}e^{i}\wedge e^{j}+a_{cb}\mathbf{e}^{c}\wedge \mathbf{e}%
^{b},  \label{hvm}
\end{gather}%
where the v-components $\check{g}_{ab}$ are def\/ined by the same
coef\/f\/icients as $\check{g}_{ij}$.

\begin{definition}
A h-v-metric on a N-anholonomic manifold is a second rank d-tensor of
type~(\ref{hvm}).
\end{definition}

We can def\/ine the local d-covector f\/ields $\eta _{i}^{i^{\prime
}}=g_{ij}\xi _{i^{\prime }}^{j}$ and the d-tensors of type $(1,1)$, $l_{j}^{i}$~and~$m_{j}^{i},$ satisfying the conditions%
\begin{gather*}
l_{j}^{i} =\xi _{i^{\prime }}^{i}\eta _{j}^{i^{\prime }} \qquad
\mbox{and} \qquad m_{j}^{i}=\delta _{j}^{i}-\xi _{i^{\prime }}^{i}\eta
_{j}^{i^{\prime }},\qquad \mbox{for}\quad i^{\prime }=1,\dots,k; \\
l_{j}^{i} =0\qquad \mbox{and}\qquad {m}_{j}^{i}=\delta _{j}^{i},\qquad \mbox{for}\quad k=0.
\end{gather*}%
For further computations, it is useful to use a matrix calculus. One denotes
$A=(a_{ij})$, $B=(b^{ij}),$ or $C=(c_{j}^{i}),$ where the index $i$ specif\/ies
the row and the index $j$ specify the column. Thus $b^{ij}a_{jk}=c_{k}^{i}$
means $BA=C$ and $a_{ij}b^{jk}=c_{i}^{k}$ means $AB={}^{t}C,$ the transpose
of~$C$. We shall consider the matrices
\begin{gather}
\widehat{g} =(g_{ij}),\qquad \widehat{a}=(a_{ij}),\qquad \widehat{\xi }=(\xi _{i^{\prime
}}^{i}),\qquad \widehat{{l}}=({l}_{j}^{i}),  \notag \\
\widehat{\eta } =(\eta _{i}^{i^{\prime }}),\qquad \widehat{{m}}=({m}_{j}^{i}),\qquad \widehat{\delta }^{\prime }=(\delta _{i^{\prime }j^{\prime }}),\qquad
\widehat{\delta }=(\delta _{j}^{i})  \label{matr01}
\end{gather}%
for which the formulas
\begin{equation*}
\widehat{a}\widehat{\xi }=0,\qquad {}^{t}\widehat{\xi }\widehat{g}\widehat{\xi }=
\widehat{\delta },\qquad ^{t}\widehat{\eta }=\widehat{g}\widehat{\xi },\qquad \widehat{%
{l}}=\widehat{\xi }\widehat{\eta },\qquad \widehat{{m}}=\widehat{%
\delta }-\widehat{{l}},
\end{equation*}%
imply%
\begin{gather}
\widehat{a}\widehat{{l}} =0,\qquad \widehat{a} \widehat{{m}}=
\widehat{a},\qquad \widehat{\xi }\widehat{\eta }=\widehat{\delta },\qquad \widehat{\eta }\widehat{{l}}=\widehat{\eta },\qquad \widehat{\eta }\widehat{{m}}=0,   \notag\\
\widehat{g}\widehat{{l}} = {}^{t}\widehat{\eta }\widehat{\eta },\qquad
\widehat{g}\widehat{{m}}=\widehat{g}- {}^{t}\widehat{\eta }\widehat{\eta },\qquad
 \widehat{g}\widehat{{l}}^{-1}= {}^{t}\widehat{\xi }\widehat{\xi },\qquad \widehat{{m}}\widehat{g}^{-1}=\widehat{g}- {}^{t}\widehat{\xi }%
\widehat{\xi }.  \label{prop01}
\end{gather}%
One follows that the matrices $\widehat{g}\widehat{{l}}$, $\widehat{g}\widehat{{m}}$, $\widehat{{l}}\widehat{g}^{-1}$, $\widehat{%
{m}}\widehat{g}^{-1}$ are symmetric. Similar matrices and formulas
can be def\/ined on the v-subspaces. They are labelled, for instance, $\widehat{g}=(g_{ab})$,
$\widehat{a}=(a_{bc})$, $\widehat{\xi }=(\xi _{a^{\prime
}}^{a})$, $\widehat{{l}}=({l}_{b}^{a})$, \dots. We shall write ${}^{h}%
\widehat{g}$ or ${}^{v}\widehat{g}$ if it would be necessary to emphasize
that a corresponding matrix is def\/ined on h- or v-subspaces. For
simplicity, we shall present only the formulas for the h-subspace and omit
similar ones for the v-subspace.

Let us denote by ${}^{d}\mathcal{X}(h\mathbf{V})$ the module of d-vector
f\/ields on $h\mathbf{V}$ (we note that $h\mathbf{V}=M$ if we take $\mathbf{V}%
^{n+n}=TM).$ One considers the submodules%
\begin{gather}
{}^{h}\mathbf{K} =\big\{\xi ^{j}\in {}^{d}\mathcal{X}(h\mathbf{V}^{n})\,|\,a_{ij}\xi ^{j}=0\big\},  \notag \\
{}^{h}\mathbf{H} =\big\{\zeta ^{i}\in {}^{d}\mathcal{X}(h\mathbf{V}^{n})\,|\,g_{ij}\xi ^{i}\zeta ^{j}=0,\ \forall\, \xi ^{i}\in \mathbf{K}\big\}.  \label{mod}
\end{gather}%
The elements of ${}^{h}\mathbf{K}$ are globally def\/ined since they are in
the kernel of the mapping $\xi ^{j}\rightarrow a_{ij}\xi ^{j}.$ The
structure ${}^{h}\mathbf{H}$ is also global because its elements are
orthogonal to ${}^{h}\mathbf{K}$ which is locally spanned by $(\xi
_{i^{\prime }}^{i}).$ One holds the following mutually equivalent conditions:%
\begin{equation}
\zeta ^{i}\in {}^{h}\mathbf{H},\qquad \eta _{i}^{i^{\prime }}\zeta ^{i}=0,\qquad
{l}_{j}^{i}\zeta ^{j}=0.  \label{eq01}
\end{equation}

One also follows:

\begin{proposition}
The system of linear equations
$a_{ij}X^{j}=0$ and $\eta _{i}^{i^{\prime }}X^{i}=0$
 has only the trivial solution $X^{j}=0.$
\end{proposition}

The formulas (\ref{prop01}) and (\ref{eq01}) result in properties
 ${l}_{j}^{i}X^{j}\in {}^{h}\mathbf{K}$ and ${m}%
_{j}^{i}X^{j}\in {}^{h}\mathbf{H} $,
for every $X^{j}\in {}^{d}\mathcal{X}(h\mathbf{V}^{n})$, and
 \begin{equation}
\widehat{{l}}+\widehat{{m}}=\widehat{\delta },\qquad \widehat{%
{l}}^{2}=\widehat{{l}},\qquad \widehat{{m}}^{2}=\widehat{%
{m}},\qquad \widehat{{l}}\widehat{{m}}=\widehat{{m}}%
\widehat{{l}}=0,\qquad {}^{t}\widehat{{l}}\widehat{g}=0,
\label{prop02}
\end{equation}%
proving that the submodules ${}^{h}\mathbf{K}$ and ${}^{h}\mathbf{H}$ are
orthogonal and supplementary, i.e.\ ${}^{d}\mathcal{X}(h\mathbf{V})={}^{h}%
\mathbf{K+}{}^{h}\mathbf{H}$, ${}^{h}\mathbf{K}\cap {}^{h}\mathbf{H=\{0\}}$
and $g_{ij}\xi ^{i}\zeta ^{j}=0,$ for every $\xi ^{i}\in {}^{h}\mathbf{K}$
and $\zeta ^{j}\in {}^{h}\mathbf{H}$. Following formulas~(\ref{prop02}), we
conclude that $\widehat{{l}}$ and $\widehat{{m}}$ are unique
projectors (tensor f\/ields) not depending on $\widehat{\xi }$ which are
completely determined, respectively, by ${}^{h}\mathbf{K}$ and ${}^{h}%
\mathbf{H}$ and globally def\/ined on $\mathbf{V}^{n+n}$.

The next step is to extend the matrix $\widehat{a}$ to a nonsingular skew
symmetric one of dimension $(n+k,n+k)$,
$
\widetilde{a}=\left[
\begin{array}{cc}
\widehat{a} & -{}^{t}\varphi \\
\varphi & 0%
\end{array}%
\right] .$
The inverse matrix $\widetilde{a}^{-1},$ satisfying the condition $%
\widetilde{a}\widetilde{a}^{-1}=\widehat{\delta },$ has the form
\begin{equation}
\widetilde{a}^{-1}=\left[
\begin{array}{cc}
\check{a} & \widehat{\xi } \\
{}^{t}\widehat{\xi } & 0%
\end{array}%
\right] ,  \label{aux00}
\end{equation}%
where the matrix $\check{a}=\left( \check{a}^{ij}\right) $ does not depend
on the choice of $\widehat{\xi }$ and it is uniquely def\/ined by $\widehat{a%
}\check{a}={}^{t}\widehat{{m}}$ and $\widehat{{l}}\check{a}%
=0,$ i.e.\ this matrix is uniquely def\/ined on $\mathbf{V}^{n+n}.$

If the nonsymmetric part of metric vanishes, $\widehat{a}=0,$ we have $%
\varphi =\widehat{\xi }^{-1}$, $\widehat{{l}}=\widehat{\delta }$, $\widehat{{m}}=0$ and $\check{a}=0.$ In the case $k=0$, we have $\widehat{%
{l}}=0$, $\widehat{{m}}=\widehat{\delta }$ and $\check{a}=%
\widehat{a}^{-1}$ and if additionally $\widehat{g}=0$, the constructions
reduces to an almost symplectic structure.

\section{Distinguished connections and nonsymmeric metrics}\label{sec3}

We consider the basic properties of linear connections adapted to the
N-connection structure on a nonholonomic manifold $\mathbf{V}^{n+n}$
enabled with nonsymmetric metric structure $\mathbf{g}$.

\subsection{Torsion and curvature of d-connections}\label{sec3.1}

In general, the concept of linear connection (adapted or not adapted to a
N-connection structure) does not depend on the concept of metric (symmetric
or nonsymmetric).

\begin{definition}
A distinguished connection (d-connection) $\mathbf{D}$ on $\mathbf{V}^{n+n}$
is a N-adapted linear connection, preserving by parallelism the vertical
and horizontal distribution (\ref{whitney}).
\end{definition}

In local form, a d-connection $\mathbf{D}=\left({}^{h}D,{}^{v}D\right) $
is given by its coef\/f\/icients $\mathbf{\Gamma }_{\ \alpha \beta }^{\gamma}=(L_{jk}^{i}, L_{bk}^{a}, C_{jc}^{i}$, $C_{bc}^{a}),$ where ${}^{h}D=(L_{jk}^{i},L_{bk}^{a})$ and ${}^{v}D=(C_{jc}^{i}, C_{bc}^{a})$ are
respectively the covariant h- and v-derivatives.

The torsion $\mathcal{T}$ of a d-connection $\mathbf{D}$, is def\/ined by the
d-tensor f\/ield
\begin{equation*}
\mathcal{T}(\mathbf{X},\mathbf{Y})\doteqdot \mathbf{D}_{\mathbf{X}}\mathbf{Y}-\mathbf{D}_{\mathbf{Y}}%
\mathbf{X}-[\mathbf{X},\mathbf{Y}],
\end{equation*}%
for any d-vectors $\mathbf{X}=h\mathbf{X}+v\mathbf{X}={}^{h}\mathbf{X}+{}^{v}\mathbf{X}$ and $\mathbf{Y}=h\mathbf{Y}+v\mathbf{Y}$, with a
corresponding N-adapted decomposition into
\begin{gather}
\mathcal{T}(\mathbf{X},\mathbf{Y}) =\{h\mathcal{T}(h\mathbf{X},h\mathbf{Y}),h%
\mathcal{T}(h\mathbf{X},v\mathbf{Y}),h\mathcal{T}(v\mathbf{X},h\mathbf{Y}),h%
\mathcal{T}(v\mathbf{X},v\mathbf{Y}),  \notag \\
\phantom{\mathcal{T}(\mathbf{X},\mathbf{Y}) =\{}{} v\mathcal{T}(h\mathbf{X},h\mathbf{Y}),v\mathcal{T}(h\mathbf{X},v\mathbf{Y}%
),v\mathcal{T}(v\mathbf{X},h\mathbf{Y}),v\mathcal{T}v\mathbf{X},v\mathbf{Y}%
)\}.  \label{tors}
\end{gather}%
The nontrivial N-adapted coef\/f\/icients of
\begin{equation*}
\mathcal{T}=\big\{\mathbf{T}_{~\beta \gamma }^{\alpha }=-\mathbf{T}_{~\gamma
\beta }^{\alpha }=\big(
T_{~jk}^{i},T_{~ja}^{i},T_{~jk}^{a},T_{~ja}^{b},T_{~ca}^{b}\big) \big\}  
\end{equation*}%
can be computed by introducing $\mathbf{X}=\mathbf{e}_{\alpha }$ and $\mathbf{Y}=\mathbf{e}_{\beta }$ into (\ref{tors}), see formulas~(\ref{dtorsc}) in Appendix and~\cite{vrfg}.

The curvature of a d-connection $\mathbf{D}$ is def\/ined
\begin{equation*}
\mathcal{R}(\mathbf{X},\mathbf{Y})\doteqdot \mathbf{D}_{\mathbf{X}}\mathbf{D}_{\mathbf{Y}}-\mathbf{D}_{\mathbf{Y}}\mathbf{D}_{\mathbf{X}}
-\mathbf{D}_{[\mathbf{X},\mathbf{Y}]},
\end{equation*}%
with N-adapted decomposition%
\begin{gather}
\mathcal{R}(\mathbf{X},\mathbf{Y})\mathbf{Z} =\{\mathcal{R}(h\mathbf{X},h\mathbf{Y})h\mathbf{Z},\ \mathcal{R}(h\mathbf{X},v\mathbf{Y})h\mathbf{Z},\ \mathcal{R}(v\mathbf{X},h%
\mathbf{Y})h\mathbf{Z},\ \mathcal{R}(v\mathbf{X},v\mathbf{Y})h\mathbf{Z},\ \notag \\
\phantom{\mathcal{R}(\mathbf{X},\mathbf{Y})\mathbf{Z} =\{}{}
\mathcal{R}(h\mathbf{X},h\mathbf{Y})v\mathbf{Z},\ \mathcal{R}(h\mathbf{X},v\mathbf{Y})v\mathbf{Z},\
\mathcal{R}(v\mathbf{X},h\mathbf{Y})v\mathbf{Z},\ \mathcal{R}(v\mathbf{X},v%
\mathbf{Y})v\mathbf{Z}\}.  \label{curv}
\end{gather}%
The formulas for local N-adapted components and their symmetries, of the
d-torsion and d-curvature, can be computed by introducing $\mathbf{X}=%
\mathbf{e}_{\alpha }$, $\mathbf{Y}=\mathbf{e}_{\beta }$ and $\mathbf{Z}=%
\mathbf{e}_{\gamma }$ in (\ref{curv}). The nontrivial N-adapted
coef\/f\/icients
\begin{equation*}
\mathcal{R}=\big\{\mathbf{R}_{\ \beta \gamma \delta }^{\alpha }=\left(
R_{\ hjk}^{i},R_{\ bjk}^{a},R_{\ hja}^{i},R_{\
bja}^{c},R_{\ hba}^{i},R_{\ bea}^{c}\right)\big\}
\end{equation*}%
are given by formulas (\ref{dcurvc}) in Appendix, see also~\cite{vrfg}.

Contracting the f\/irst and forth indices $\mathbf{R}_{\ \beta \gamma
}=\mathbf{R}_{\ \beta \gamma \alpha }^{\alpha }$, one gets the N-adapted
coef\/f\/icients for the Ricci tensor%
\begin{equation*}
\mathrm{Ric} \doteqdot \{\mathbf{R}_{\beta \gamma }= \left(
R_{ij},R_{ia},R_{ai},R_{ab}\right) \}, 
\end{equation*}%
see formulas (\ref{driccic}) in Appendix and~\cite{vrfg}. It should be
noted here that for general d-connections the Ricci tensor is not
symmetric, i.e.\ $\mathbf{R}_{\beta \gamma }\neq \mathbf{R}_{\gamma
\beta }.$

On spaces of dimension $n+n,$ it is convenient to work with a particular
class of d-connections.

\begin{definition}
\label{defncdc}A normal d-connection ${}_{n}\mathbf{D}$ is compatible with
the almost complex structu\-re~$\mathbb{F}$~(\ref{acs}), i.e.\ satisf\/ies the
condition
\begin{equation}
{}_{n}\mathbf{D}_{\mathbf{X}}\mathbb{F}=0,  \label{acscomp}
\end{equation}%
for any d-vector $\mathbf{X}$ on $\mathbf{V}^{n+n}.$
\end{definition}

From local formulas for (\ref{acscomp}), one follows:

\begin{theorem}
A normal d-connection is characterized by a pair of local coefficients ${}_{n}\mathbf{\Gamma }_{\ \alpha \beta }^{\gamma }=\big({} _{n}L_{jk}^{i},{}_{n}C_{bc}^{a}\big) $ defined by conditions
\begin{gather*}
{}_{n}\mathbf{D}_{\mathbf{e}_{k}}(\mathbf{e}_{j})  = {}_{n}L_{jk}^{i}%
\mathbf{e}_{i},\qquad {}_{n}\mathbf{D}_{\mathbf{e}_{k}}(e_{a})={}_{n}L_{ak}^{b}e_{b}:\qquad
 \mbox{for} \quad j =a,\ i=b,\qquad {}_{n}L_{jk}^{i}={} _{n}L_{ak}^{b}, \\
{}_{n}\mathbf{D}_{e_{c}}(\mathbf{e}_{j}) ={} _{n}C_{jc}^{i}\mathbf{e}_{i},\qquad {}_{n}\mathbf{D}_{e_{c}}(e_{a})={}_{n}C_{ac}^{b}e_{b}: \qquad  \mbox{for}\quad j =a,\ i=b,\qquad {}_{n}C_{jc}^{i}={}_{n}C_{ac}^{b}.
\end{gather*}
\end{theorem}

By a straightforward local calculus introducing the pairs of coef\/f\/icients for%
${}_{n}\mathbf{\Gamma }_{\ \alpha \beta }^{\gamma },$ respectively, in (%
\ref{tors}) and (\ref{curv}), see also formulas (\ref{dtorsc}) and (\ref{dcurvc}) in Appendix, we prove

\begin{corollary}
The N-adapted coefficients of torsion ${}_{n}\mathcal{T} =\{{}_{n}T_{\
\beta \gamma }^{\alpha }\}$ and curvature ${}_{n}\mathcal{R}=\{{}_{n}R_{\
\beta \gamma \tau }^{\alpha }\}$ of the normal d-connection ${}_{n}\mathbf{D}$ are defined respectively by formulas
\begin{gather*}
{}_{n}T_{\ jk}^{i} ={}_{n}L_{\ jk}^{i}-{}_{n}L_{\ kj}^{i},\qquad {} _{n}T_{\
ja}^{i}={}_{n}C_{\ ja}^{i},\qquad {}_{n}T_{\ ji}^{a}=\Omega _{\ ji}^{a},  \\
{}_{n}T_{\ bi}^{a} =\frac{\partial N_{i}^{a}}{\partial y^{b}}-{}_{n}L_{\
bi}^{a},\qquad {}_{n}T_{\ bc}^{a}={}_{n}C_{\ bc}^{a}-{}_{n}C_{\ cb}^{a},
\end{gather*}%
and
\begin{gather*}
{}_{n}R_{\ hjk}^{i}  = \mathbf{e}_{k}\left({}_{n}L_{\ hj}^{i}\right) -%
\mathbf{e}_{j}\left({}_{n}L_{\ hk}^{i}\right) +{}_{n}L_{\ hj}^{m}{}_{n}L_{\ mk}^{i}-{}_{n}L_{\ hk}^{m}{}_{n}L_{\
mj}^{i}-{}_{n}C_{\ ha}^{i}\Omega _{\ kj}^{a}, \\
{}_{n}R_{\ bka}^{c}  = e_{a}\left({} _{n}L_{\ bk}^{c}\right) -{}_{n}%
\mathbf{D}_{k}  \left({}_{n}C_{\ ba}^{c}\right) +{}_{n}C_{\ bd}^{c}{}_{n}T_{\ ka}^{c}, \\
{}_{n}R_{\ bcd}^{a}  = e_{d} \left({}_{n}C_{\ bc}^{a}\right) -e_{c} \left({}_{n}C_{\ bd}^{a}\right) +{}_{n}C_{\ bc}^{e}{}_{n}C_{\ ed}^{a}-{}_{n}C_{\ bd}^{e}{}_{n}C_{\ ec}^{a}.
\end{gather*}
\end{corollary}

In the geometry of N-anholonomic manifolds and Finsler--Lagrange spa\-ces
an important role is given to a special type of d-connections:

\begin{definition}
A normal d-connection ${}_{c}\mathbf{\Gamma }_{\ \alpha \beta }^{\gamma
}=\left({}_{c}L_{jk}^{i},{}_{c}C_{bc}^{a}\right) $ on $\mathbf{V}^{n+n}$ is
a Cartan d-connection if it satisf\/ies the conditions%
\begin{equation}
{}_{c}^{h}D_{k}y^{a}=0\qquad \mbox{and}\qquad {}_{c}^{v}D_{a}y^{b}=\delta _{a}^{b}.
\label{ccd}
\end{equation}
\end{definition}

By an explicit local N-adapted calculus, we can verify:

\begin{proposition}
The N-connection and Cartan d-connection coefficients satisfy the
conditions%
\begin{equation*}
N_{i}^{a}=y^{b}{}_{c}L_{bi}^{a}\qquad \mbox{and}\qquad y^{b}{}_{c}C_{bc}^{a}=0
\end{equation*}%
and the d-torsions and d-curvatures are related by formulas%
\begin{equation*}
{}_{c}T_{\ kj}^{a}=\Omega _{\ kj}^{a}=y^{b}{}_{c}R_{\ bjk}^{a},\qquad {}_{c}T_{\
bi}^{c}=y^{a}{}_{c}R_{\ akb}^{c},\qquad {}_{c}T_{\ bc}^{a}=y^{d}{}_{c}R_{\ dbc}^{a}.
\end{equation*}
\end{proposition}

Finally, we note that on N-anholonomic manifolds we can work equivalently
with dif\/ferent classes of d-connections. For applications in modern physics~\cite{vrfg,vncg,vrf02,avnsm01}, the constructions with metric compatible
d-connections which are def\/ined in a unique way by a metric and certain
prescribed torsion structures are considered to be more related to standard
theories.

\subsection{Distinguished connections compatible
with nonsymmetric metrics}\label{sec3.2}

In this section, we def\/ine a class of d-connections which are compatible
with nonsymmetric metric structures.

Let us consider a nonholonomic manifold $\mathbf{V}^{n+n}$ with f\/ixed
N-connection $\mathbf{N}$ and enabled with a d-connection $\mathbf{D}_{\alpha }=(D_{i},D_{a})$ and nonsymmetric metric structure $\check{\mathbf{g}}=\mathbf{g}+\mathbf{a.}$ A nonsymmetric metric $\check{\mathbf{g}}$ is
characterized by d-tensor f\/ields/matrices (\ref{matr01}) satisfying the
properties~(\ref{prop01}) and~(\ref{prop02}). Considering actions of a
general d-connection $\mathbf{D}$ on the mentioned formulas, one prove:

\begin{proposition}
One holds the following formulas for the h-covariant (v-covariant)
derivatives of the coefficients of matrices \eqref{matr01} defined globally
by a nonsymmetric metric structure:
\begin{gather}
{l}_{p}^{i}  {l}_{j}^{q}  D_{k}{l}_{q}^{p} =0,\qquad
{m}_{p}^{i}D_{k}{l}_{q}^{p}=\ {l}_{q}^{s}\ D_{k}{%
l}_{s}^{i},\qquad {m}_{j}^{s}D_{k}{l}_{s}^{i}=\ {l}_{q}^{i}\
D_{k}{l}_{j}^{q},  \notag \\
{l}_{i}^{p}  {l}_{j}^{q}  D_{k}a_{pq}  = 0,\qquad a_{sj}\check{a}%
^{ir} D_{k}{l}_{r}^{s}=0,\qquad {l}_{j}^{s}(\check{a}^{ir}D_{k}a_{rs}-D_{k}{l}_{s}^{i})=0  \label{aux03}
\end{gather}%
(one v-subspace the formulas are similar but with ${}^{h}D=\{D_{k}\}\rightarrow {}^{v}D=\{D_{a}\}).$
\end{proposition}

\begin{proof}
In order to prove the formulas (\ref{aux03}), it is convenient to use the
matrix covariant h-derivative $D_{k}\widehat{{l}}=\left( D_{k}
{l}_{j}^{i}\right) $ and v-derivative $D_{a}\widehat{{l}}
=\left( D_{a}{l}_{j}^{i}\right) $ which for $\widehat{{l}}^{2}=
\widehat{{l}}$ imply respectively $D_{k}\widehat{{l}}
\widehat{{l}}+\widehat{{l}} D_{k}\widehat{{l}}=
\widehat{{l}}$ and $D_{a}\widehat{{l}} \widehat{{l}}+
\widehat{{l}} D_{a}\widehat{{l}}=\widehat{{l}}.$ One
should be used the matrix relation $\widehat{a}\widehat{{l}}=0$ to
get the formulas containing the anti-symmetric part of metric. \end{proof}

We shall work with a more restricted class of d-connections on $\mathbf{V}%
^{n+n}$:

\begin{definition}
A d-connection $\mathbf{D}=\{\mathbf{\Gamma} _{\ \alpha \beta }^{\gamma
}=\left( L_{jk}^{i},C_{bc}^{a}\right) \}$ is compatible with a
nonsymmetric d-metric $\check{\mathbf{g}}$ if
\begin{equation}
\mathbf{D}_{k}\check{g}_{ij}=0\qquad \mbox{and}\qquad \mathbf{D}_{a}\check{g}_{ij}=0.
\label{nscomp}
\end{equation}
\end{definition}

For the d-metric (\ref{hvm}), the equations (\ref{nscomp}) are written
 $
\mathbf{D}_{k}g_{ij}=0$, $\mathbf{D}_{a}g_{bc}=0$, $\mathbf{D}_{k}a_{ij}=0$, $\mathbf{D}_{e}a_{bc}=0.$

In such cases, there are additional to (\ref{aux03}) properties:

\begin{proposition}
\label{prmc}The d-tensor fields $\widehat{{l}}=({l}_{j}^{i})$,
$\widehat{{m}}=({m}_{j}^{i})$, $\check{g}=(g^{ij})$ and $\check{a}%
=\left( \check{a}^{ij}\right) $ satisfy the conditions%
\begin{gather*}
D_{k}{l}_{j}^{i} =0,\qquad D_{k}{m}_{j}^{i}=0,\qquad D_{k}g^{ij}=0,\qquad
D_{k}\check{a}^{ij}=0, \\
D_{a}{l}_{c}^{b}  = 0,\qquad D_{a}{m}_{c}^{b}=0,\qquad D_{a}g^{bc}=0,\qquad
D_{a}\check{a}^{bc}=0,
\end{gather*}%
for any $\mathbf{D}$ compatible to $\check{\mathbf{g}.}$
\end{proposition}

\begin{proof}
It is important to take the covariant h- and v-derivatives for formulas $%
a_{ij}\xi _{i^{\prime }}^{i}=0,$ $g_{ij}\xi _{i^{\prime }}^{i}\xi
_{j^{\prime }}^{j}=\delta _{i^{\prime }j^{\prime }}$, $a_{rs}$ $\check{a}%
^{sj}={m}_{r}^{j}$ and $l_{s}^{i}$ $\check{a}^{sj}=0$ and use the
properties (\ref{prop01}) and (\ref{prop02}).
\end{proof}

It is important to def\/ine the geometric properties of tensor f\/ields from
Proposition~\ref{prmc} because they are used to formulate and prove main
Theorems~\ref{thau4} and~\ref{thau5}.

To work with metric compatible connections is not only a preferred approach
in order to elaborate more ``simple'' physical theories, but it is motivated
geometrically by the fact that there is a method of metrization for
arbitrary d-connections.

\section{(Non)symmetric metrization procedure of d-connections}\label{sec4}

Usually, in gravity theories, one f\/ix a linear connection connection
structure which is, or not, metric compatible and, in general, with
nontrivial torsion (in general relativity, this is the Levi-Civita
connection which by def\/inition is both metric compatible and torsionless).
In Finsler geometry, A.~Kawaguchi \cite{kaw1,kaw2,kaw3} proposed the method
of metrization of d-connections which was further developed for Lagrange
spaces, see~\cite{ma,ma1987}. The approach was also used for Einstein and
Riemann--Cartan spaces def\/ined by generic of\/f-diagonal metrics when the
nonholonomic deformations were considered not only for frame and metric
structures but also for linear connections in order to generate such ansatz
for geometric objects when the f\/ield equations became exactly integrable for
certain systems on nonlinear partial dif\/ferential equations, see review~\cite{vrfg}. It was emphasized that imposing additional constraints on integral
varieties of solutions for more general classes of connections it is
possible to generate solutions for the Levi-Civita connection. In this
section, we show how the metrization method can be applied on N-anholonomic
manifolds enabled with nonsymmetric metrics.

A f\/ixed metric structure on a nonholonomic manifold induces certain (Obata)
type operators def\/ining the set of d-connections being compatible with this
metric structure. The aim of this section is to def\/ine the properties of
such operators.

\subsection{Metrization methods with symmetric metrics}\label{sec4.1}

By straightforward computations, for symmetric metrics on vector bundles,
one proved two important results which hold true for N-anholonomic
manifolds:

\begin{enumerate}\itemsep=0pt
\item \textit{Kawaguchi's metrization:} To any f\/ixed d-connection ${}_{\circ }\mathbf{D}$ we can associate a compatible with a metric $\mathbf{g}
$ d-connection $\mathbf{D}$, satisfying the condition $\mathbf{D}_{\mathbf{X}}\mathbf{g=0}$ for any d-vector on~$\mathbf{V}^{n+n}$. The N-adapted
coef\/f\/icients for d-connections are related by formulas%
\begin{gather*}
L_{\ jk}^{i} ={}_{\circ }L_{\ jk}^{i}+\frac{1}{2}g^{im}{}_{\circ
}D_{k}g_{mj},\qquad L_{\ bk}^{a}={}_{\circ }L_{\ bk}^{a}+\frac{1}{2}g^{ac}{}_{\circ }D_{k}g_{cb}, \\
C_{\ jc}^{i} ={}_{\circ }C_{\ jc}^{i}+\frac{1}{2}g^{im}{}_{\circ
}D_{c}g_{mj},\qquad C_{\ bc}^{a}={}_{\circ }L_{\ bc}^{a}+\frac{1}{2}g^{ae}{}_{\circ }D_{c}g_{eb}.
\end{gather*}

\item \textit{Miron's procedure:} The set of d-connections $\{\mathbf{D}%
\} $ satisfying the conditions $\mathbf{D}_{\mathbf{X}}\mathbf{g=0}$ for a~given $\mathbf{g}$ is def\/ined by formulas%
\begin{gather*}
L_{\ jk}^{i} = \widehat{L}_{jk}^{i}+{}^{-}O_{km}^{ei}\mathbf{X}%
_{ej}^{m},\qquad L_{\ bk}^{a}=\widehat{L}_{bk}^{a}+{}^{-}O_{bd}^{ca}\mathbf{Y}%
_{ck}^{d}, \\
C_{\ jc}^{i} = \widehat{C}_{jc}^{i}+{}^{+}O_{jk}^{mi}\mathbf{X}%
_{mc}^{k},\qquad C_{\ bc}^{a}=\widehat{C}_{bc}^{a}+{}^{+}O_{bd}^{ea}\mathbf{Y}
_{ec}^{d},
\end{gather*}%
where
\begin{equation}
{}^{\pm }O_{jk}^{ih}=\frac{1}{2}(\delta _{j}^{i}\delta _{k}^{h}\pm
g_{jk}g^{ih}),\qquad {}^{\pm }O_{bd}^{ca}=\frac{1}{2}(\delta _{b}^{c}\delta
_{d}^{a}\pm g_{bd}g^{ca})  \label{obop}
\end{equation}%
are the so-called the Obata operators; $\mathbf{X}_{ej}^{m}$, $\mathbf{X}_{mc}^{k}$,
$\mathbf{Y}_{ck}^{d}$ and $\mathbf{Y}_{ec}^{d}$ are arbitrary
d-tensor f\/ields and $\widehat{\mathbf{\Gamma }}_{\ \alpha \beta }^{\gamma
}=\big( \widehat{L}_{jk}^{i},\widehat{L}_{bk}^{a},\widehat{C}_{jc}^{i},%
\widehat{C}_{bc}^{a}\big) $, with
\begin{gather}
\widehat{L}_{jk}^{i} =\frac{1}{2}g^{ir}\left(
e_{k}g_{jr}+e_{j}g_{kr}-e_{r}g_{jk}\right) , \notag \\
\widehat{L}_{bk}^{a} =e_{b}(N_{k}^{a})+\frac{1}{2}g^{ac}\big(
e_{k}g_{bc}-g_{dc}\ e_{b}N_{k}^{d}-g_{db}\ e_{c}N_{k}^{d}\big) ,  \notag \\
\widehat{C}_{jc}^{i} =\frac{1}{2}g^{ik}e_{c}g_{jk},\qquad \widehat{C}_{bc}^{a}=
\frac{1}{2}g^{ad}\left( e_{c}g_{bd}+e_{c}g_{cd}-e_{d}g_{bc}\right)   \label{candcon}
\end{gather}%
is the canonical d-connections uniquely def\/ined by the coef\/f\/icients of
d-metric $\mathbf{g=}[g_{ij},g_{ab}]$ and N-connection $\mathbf{N}%
=\{N_{i}^{a}\}$ in order to satisfy the conditions $\widehat{\mathbf{D}}_{\mathbf{X}}\mathbf{g}=\mathbf{0}$ and $\widehat{T}_{\ jk}^{i}=0$ and $\widehat{T}_{\
bc}^{a}=0$ but $\widehat{T}_{\ ja}^{i}$, $\widehat{T}_{\ ji}^{a}$ and $\widehat{%
T}_{\ bi}^{a}$ are not zero, see formulas (\ref{dtorsc}). On $\mathbf{V}%
^{n+n}$, it is possible to work with the normal, or the Cartan,
d-connection as we def\/ined by (\ref{acscomp}), or (\ref{ccd}).
\end{enumerate}

\subsection{Properties of Obata operators for nonsymmetric d-metrics}\label{sec4.2}

The A.~Kawaguchi and R.~Miron metrization procedures can be generalized for
nonsymmetric metrics $\check{\mathbf{g}}=\mathbf{g}+\mathbf{a}$ (\ref{hvm})
on N-anholonomic manifolds. For simplicity, we consider the formulas for
h-subspaces of a $\mathbf{V}^{n+n}$ with f\/ixed N-connection structure.

On h-subspace, for the symmetric part $\mathbf{g}$ of $\check{\mathbf{g}}$,
the Obata operators are of type ${}_{h}^{[a]}O=\{{}^{\pm }O_{jk}^{ih}\}$ (\ref{obop}), for $[a]=\pm .$ In general, we have on $\mathbf{V}^{n+n}$ the
operators $\mathbf{O}=\left({}_{h}^{[a]}O,{}_{v}^{[a]}O\right) .$ These
operators act on d-tensor f\/ields of type $(1,2)$ following the rule%
$\left({}_{h}^{\pm }OX\right) _{jk}^{i}={}^{\pm }O_{sj}^{ih} \mathbf{X}_{hk}^{s},$
when the product of Obata h-operators, ${}_{h}^{[a]}O{}_{h}^{[b]}O,$ is
def\/ined%
\begin{equation}
\left({}_{h}^{[a]}O{}_{h}^{[b]}O\right) _{sj}^{im}={}_{h}^{[a]}O_{kj}^{ir}{}_{h}^{[b]}O_{sr}^{km}.  \label{aux4}
\end{equation}%
We can check by explicit computations:

\begin{proposition}
\label{posp}There are satisfied the properties
\begin{equation*}
{}_{h}^{-}O+{}_{h}^{+}O=I,\qquad \big({}_{h}^{[a]}O\big) ^{2}={}_{h}^{[a]}O,\qquad
{}_{h}^{-}O{}_{h}^{+}O={}_{h}^{+}O{}_{h}^{-}O=0,  \notag
\end{equation*}%
where $I$ is the identity matrix $\delta _{s}^{i}\delta _{j}^{r}$, when $I\mathbf{X}=\mathbf{X}$.
\end{proposition}

This Proposition states that the operators $\mathbf{O}$ are supplementary
projectors on the module of d-tensor f\/ields of type $(1,2)$.

The skewsymmetric part $\mathbf{a}$ of $\check{\mathbf{g}}$ def\/ines an
additional set of Obata operators $\mathbf{\Phi }=\big({}_{h}^{[a]}\Phi ,{}_{v}^{[a]}\Phi \big) $, where, for instance,
 $
2\left({}_{h}^{\pm }\Phi \right) _{sj}^{ir}=\delta _{s}^{i}\delta
_{j}^{r}\pm \big( {l}_{s}^{i}{l}_{j}^{r}+a_{sj}\check{a}
^{ir}\big) $.

Introducing the operator ${}_{h}\theta =\{\theta _{sj}^{ir}\}$, with
 $\theta _{sj}^{ir}=\frac{1}{2}\big( {l}_{s}^{i} {m}_{j}^{r}+
{m}_{s}^{i}\ {l}_{j}^{r}\big) $,
one proves by algebraic computations:

\begin{proposition}
One holds the relations
\begin{equation}
{}_{h}^{-}\Phi +{}_{h}^{+}\Phi =I,\qquad \big({}_{h}^{[a]}\Phi \big) ^{2}={}_{h}^{[a]}\Phi -\frac{1}{2}
{}_{h}\theta ,\qquad {}_{h}^{-}\Phi {}_{h}^{+}\Phi ={}_{h}^{+}\Phi {}_{h}^{-}\Phi =\frac{1}{2}{}_{h}\theta .  \label{aux5}
\end{equation}
\end{proposition}

The relations for ``skewsymmetric'' Obata operators (\ref{aux5}) are
dif\/ferent from those for the symmetric ones (\ref{aux4}). We can modify the
constructions by introducing $\mathbf{\Psi }=\big({}_{h}^{[a]}\Psi ,{}_{v}^{[a]}\Psi \big) ,$ where, for instance,
 ${}_{h}^{\pm }\Psi ={}_{h}^{\pm }\Phi \pm {}_{h}\theta .$

By direct computations, one proves

\begin{proposition}
\label{pons}The operators $\mathbf{\Psi }$ are supplementary projectors on
the module of d-tensor fields of type $(1,2)$ and satisfy the conditions
(for simplicity, we state them for the h-subspace):
\begin{gather}
 {}_{h}^{-}\Psi +{}_{h}^{+}\Psi =I,\qquad \big({}_{h}^{[a]}\Psi \big) ^{2}={}_{h}^{[a]}\Psi ,\qquad {}_{h}^{-}\Psi{}_{h}^{+}\Psi ={}_{h}^{+}\Psi{} _{h}^{-}\Psi
=0, \qquad {l}_{s}^{i} a_{sj}\big({}_{h}^{\pm }\Psi \big) _{pr}^{sm}=0, \notag \\
{l}_{s}^{i}{m}_{j}^{r}\big({}_{h}^{\pm }\Psi \big) _{pr}^{sm}=0,\qquad
 {}_{h}^{+}\Psi{}_{h}\theta ={}_{h}\theta {}_{h}^{+}\Psi ,\qquad {}_{h}^{-}\Psi {}_{h}\theta ={}_{h}\theta{}_{h}^{-}\Psi ={}_{h}\theta .  \label{aux6}
\end{gather}
\end{proposition}

One follows from properties (\ref{aux5}) and (\ref{aux6}) that the operators
${}_{h}^{[a]}\Phi $ and ${}_{h}^{[a]}\Psi $ commute with~$\theta $.

Finally, we note that even a number of properties of Obata operators ${}_{h}^{[a]}\Psi $ (or ${}_{h}^{[a]}\Phi )$, for the skewsymmetric part $\mathbf{a}$ of a nonsymmetric metric $\check{\mathbf{g}}=\mathbf{g}+\mathbf{a}$ are similar to the properties of Obata operators ${}_{h}^{[a]}O$ for the
symmetric part $\mathbf{g}$, there is a substantial dif\/ference between these
two classes of projectors. If $\theta $ commutes with ${}_{h}^{[a]}O,$ the
operators ${}_{h}^{[a]}\Phi $ and ${}_{h}^{[a]}\Psi $ do not necessarily
commute with ${}_{h}^{[a]}O$. For instance, we have
\begin{equation}
\left({}_{h}^{+}O{}_{h}^{+}\Phi -{}_{h}^{+}\Phi{} _{h}^{+}O\right)
_{pj}^{im}=\frac{1}{4}\left( g_{sj}g^{ir}a_{pr}\check{a}^{sm}-a_{sj}\check{a}%
^{ir}g_{pr}g^{sm}\right) .  \label{aux7}
\end{equation}
We need some additional suppositions on commutations of symmetric and
skewsymmetric Obata operators imposing certain nonholonomic constraints on
the components of nonsymmetric metrics.

\subsection{Natural nonsymmetric metrics}\label{sec4.3}

Let us def\/ine a subclass of nonsymmetric d-metrics for which the procedure
of metrization of d-connections will be commutative both for the symmetric
and skewsymmetric parts.

\begin{definition}
A nonsymmetric d-metric $\check{\mathbf{g}}=\mathbf{g}+\mathbf{a}$ is
natural (\ref{hvm}) if its associated symmetric and skewsymmetric Obata
operators satisfy at least one of the conditions (for simplicity, we state
them for the h-part)%
\begin{equation}
{}_{h}^{[a]}O{}_{h}^{[b]}\Phi ={}_{h}^{[a]}\Phi{}_{h}^{[b]}O\qquad \mbox{and}\qquad
{}_{h}^{[a]}O{}_{h}^{[b]}\Psi ={}_{h}^{[a]}\Psi {}_{h}^{[b]}O.  \label{obcr}
\end{equation}
\end{definition}

We emphasize that, if one of the commutation rules holds true from the set
of eight ones~(\ref{obcr}), the statement is true for all projectors. So, we
can verify the commutation rule only for any two operators and consider that
similar commutation rules exist for other operators.

Let us consider a set of unknown d-tensor of type $(1,2)$ denoted by
symbols $\mathbf{Y}$, $\mathbf{U}$, $\mathbf{V}$, $\mathbf{W}$ which are supposed to be connected to actions of
Obata operators on $\mathbf{X}$ being also a d-tensor of type $(1,2).$ One
holds:

\begin{proposition}
\label{prsoleq}The equation ${}_{h}^{-}O\mathbf{X}=0$ has the general
solution $\mathbf{X}={}_{h}^{+}O\mathbf{Y}$, where $\mathbf{Y}$ is
arbitrary. There are also mutually equivalent the equations ${}_{h}^{-}\Phi
\mathbf{X}=\mathbf{W}$ and ${}_{h}^{-}\Psi \mathbf{X}=\mathbf{W}+2{}_{h}\theta \mathbf{W}$.
\end{proposition}

\begin{proof}
The statement on general solution is an obvious conclusion from the
properties of matrix equations. The statement on equivalence of two matrix
equations follows from the fact that multiplying those equations on $\theta $
one gets $\theta \mathbf{X}=2 \theta \mathbf{W}$ which proves the
equivalence.
\end{proof}

For the Obata operators def\/ining the class of natural nonsymmetric
d-metrics, we provide:

\begin{theorem}
Let us suppose that ${}_{h}^{+}O{}_{h}^{+}\Psi ={}_{h}^{+}\Psi {}_{h}^{+}O$
and consider the system of equations%
\begin{equation}
{}_{h}^{-}O\mathbf{X}=\mathbf{U},\qquad {}_{h}^{+}\Psi  \mathbf{X}=\mathbf{V}
\label{eqaux}
\end{equation}%
has solutions if and only if
\begin{equation}
{}_{h}^{+}O\mathbf{U}=0,\qquad {}_{h}^{+}\Psi \mathbf{V}=0,\qquad
{}_{h}^{-}\Psi \mathbf{U}={}_{h}^{-}O\mathbf{V}.  \label{condaux}
\end{equation}%
The general solutions \eqref{eqaux} can be written in two equivalent forms%
\begin{equation}
\mathbf{X} = \mathbf{U}+{}_{h}^{+}O(\mathbf{V}+{}_{h}^{+}\Psi \mathbf{Y})
= \mathbf{V}+{}_{h}^{+}\Psi (\mathbf{U}+{}_{h}^{+}O\mathbf{Y}),  \label{soleqaux}
\end{equation}%
for arbitrary d-tensors $\mathbf{Y}$.
\end{theorem}

\begin{proof}
The necessity conditions follow from Propositions~\ref{posp} and~\ref{pons}
and conditions~(\ref{obcr}). To prove the enough conditions we assume that
the conditions (\ref{condaux}) are satisf\/ied. The f\/irst equation in (\ref{eqaux}) is equivalent to ${}_{h}^{-}O(\mathbf{X}-\mathbf{U})=0$ having the
general solution $\mathbf{X}=\mathbf{U}+{}_{h}^{+}O\mathbf{Y}$, with
arbitrary $\mathbf{Y}$, which solves also the equation ${}_{h}^{-}\Psi
\mathbf{X}=\mathbf{V}$ if and only if ${}_{h}^{-}\Psi (\mathbf{U}+{}_{h}^{+}O%
\mathbf{Y})=\mathbf{V}$. We have ${}_{h}^{-}\Psi =I-{}_{h}^{+}\Psi $ which
allows to write $\mathbf{U}+{}_{h}^{+}O\mathbf{Y}=\mathbf{V}+{}_{h}^{+}\Psi (\mathbf{}U+{}_{h}^{+}O\mathbf{Y})$ proving that the d-tensor $\mathbf{X}$
can be represented in the forms~(\ref{soleqaux}) when obviously both forms
solve the equations~(\ref{eqaux}).
\end{proof}

We can formulate a criteria of existence of normal nonsymmetric d-metrics:

\begin{theorem}
\label{thsf1}A nonsymmetric d-metric $\check{g}_{ij}(x,y)$ on
N-anholonomic manifold $\mathbf{V}^{n+n}$ is natural if and only if one
exists a non-vanishing real scalar function $\nu (x,y)$ such that
\begin{equation}
\check{a}_{ij}=\nu a_{ij}  \label{aux9}
\end{equation}%
where $a_{ij}=g_{ir}g_{js}\check{a}^{rs}.$
\end{theorem}

\begin{proof}
Taking from the conditions~(\ref{obcr}) the relation ${}_{h}^{+}O{}_{h}^{+}\Psi ={}_{h}^{+}\Psi {}_{h}^{+}O,$ we express (\ref{aux7}),
mul\-tip\-lied on $g_{im}g_{th},$ in the form
\begin{equation}
\check{a}_{ij}a_{kl}=\check{a}_{kl}a_{ij}.  \label{aux8}
\end{equation}%
Contracting this expression by $\check{a}^{lk}$ together with $a_{lk}\check{a}^{lk}={m}_{k}^{k}=n-k\neq 0$, we get (\ref{aux9}) with $\nu =(\check{a}_{lk}\check{a}^{lk})/n-k\neq 0$ because otherwise we get $\check{a}_{lk}=0$
and $\check{a}^{lk}=0$ imposing ${m}_{j}^{i}=0$ which contradicts $%
k<n.$ Conversely, the equation~(\ref{aux9}) with $\nu \neq 0$ transforms (\ref{aux8}) into ${}_{h}^{+}O{}_{h}^{+}\Psi ={}_{h}^{+}\Psi {}_{h}^{+}O$.
\end{proof}

Two examples how to construct natural nonsymmetric metrics in explicit form
will be provided below in Section~\ref{ssensm}. Finally, we emphasize that
the metrization method proposed in this section is more particular than the
Kawaguchi and Miron methods outlined in~\cite{ma}, although it is in a more
general setting of Eisenhart N-anholonomic manifolds.

\section{d-connections compatible with nonsymmetric metrics}\label{sec5}

For any natural nonsymmetric d-metric, we can def\/ine in explicit form the
set of d-connections compatible to this metric structure.The goal of this
section is to provide explicit methods of constructing natural nonsymmetric
d-metrics and elaborate a method for metrization of arbitrary
d-connections, with respect to such a d-metric.

\subsection{Examples of natural nonsymmetric metrics}
\label{ssensm}

Let us consider two quadruplets $\big( g_{ij},{}_{\pm
}F_{j}^{i},\xi _{i^{\prime }}^{j},\eta _{i}^{i^{\prime }}\big) $, where $g_{ij}$ is a symmetric
d-metric; ${}_{+}\widehat{F}=({}_{+}F_{j}^{i})$ and ${}_{-}\widehat{F}=({}_{-}F_{j}^{i})$ are d-tensor f\/ields; $\xi _{i^{\prime
}}^{j}$ are $k$ d-covector f\/ields and $\eta _{i}^{i^{\prime }}=g_{ij}\xi
_{i^{\prime }}^{j},$ where $i,j,\ldots=1,2,\dots,n$ and $i^{\prime },j^{\prime
},\ldots=1,2,\dots,k.$

\begin{definition}
A respective quadruplet def\/ines a $\big( \widehat{g},{}_{\pm }\widehat{F},
\widehat{\xi },\widehat{\eta }\big) $-structure of index $k$ if one holds
the conditions%
\begin{equation}
{}_{\pm }\widehat{F}^{2}=\mp \widehat{\delta }\pm \widehat{\xi }\widehat{
\eta },\qquad \widehat{\eta }{}_{\pm }\widehat{F}=0,\qquad
{}_{\pm }\widehat{F}\widehat{\xi }=0,\qquad \widehat{\eta }\widehat{\xi }=\widehat{\delta },\qquad
{}_{\pm }^{t}\widehat{F}\widehat{g}{}_{\pm }\widehat{F}=\pm \widehat{g}\mp {}^{t}
\widehat{\eta } \widehat{\eta }.  \label{con03}
\end{equation}
\end{definition}

For $k>0,$ one holds $\big({}_{\pm }\widehat{F}\big) ^{3}={}_{\pm }%
\widehat{F}.$ One assumes that for $k=0$ the conditions (\ref{con03})
transform into ${}_{\pm }\widehat{F}^{2}=\mp \widehat{\delta }$ and ${}_{\pm
}^{t}\widehat{F}\widehat{g}{}_{\pm }\widehat{F}=\pm \widehat{g}$ and def\/ine
an almost Hermitian structure for ${}_{+}\widehat{F}$, or an almost
hyperbolic structure for ${}_{-}\widehat{F}$. In the hyperbolic case, there is
an eigen d-vector $v$ such that $_{-}\widehat{F}Z=\pm Z$, when ${}_{-}^{t}%
\widehat{F}\widehat{g}{}_{-}\widehat{F}=-\widehat{g}$ leads to ${}^{t}Z
\widehat{g}Z=0,$ which holds for a nonpositively def\/ined symmetric metric $\widehat{g}$.

\begin{theorem}
\label{thsf2}We obtain a natural nonsymmetric d-metric $\check{g}%
_{ij}=g_{ij}+a_{ij}$ of index $k,$ with $\widehat{a}=c^{-1}\widehat{g}{}_{\pm }\widehat{F}$ and respective nonvanishing on $\mathbf{V}^{n+n}$ scalar
function $\nu =\mp c^{2}$ for a given $\big( \widehat{g},{}_{\pm }\widehat{F%
},\widehat{\xi },\widehat{\eta }\big) $-structure of index~$k.$
\end{theorem}

\begin{proof}
By straightforward computations, we can verify that a sum $g_{ij}+a_{ij}$,
where the symmetric and skew-symmetric parts are connected to the $\big(
\widehat{g},{}_{\pm }\widehat{F},\widehat{\xi },\widehat{\eta }\big) $-structure of index $k$ and scalar function in the theorem, is a natural nonsymmetric metric.
\end{proof}

The inverse statement also holds true. To show this we note that for modules~(\ref{mod}) one follows that $ \widehat{g}\widehat{{m}}={}^{t}%
\widehat{{m}}\widehat{g}$ which proves:

\begin{theorem}
\label{thsf3}For any natural nonsymmetric metric $\check{g}
_{ij}=g_{ij}+a_{ij}$ of index $k$ with $\nu =\mp c^{2},$ we can define
values the ${}_{\pm }\widehat{F}^{2}=\mp c\widehat{g}^{-1}\widehat{a}$ (or,
equivalently, ${}_{\pm }\widehat{F}^{2}=\mp c^{-1}\check{a}\widehat{g})$
which for the quadruplet $\big( g_{ij},{}_{\pm }F_{j}^{i},\xi _{i^{\prime
}}^{j},\eta _{i}^{i^{\prime }}\big) $ states a $\big( \widehat{g},{}_{\pm
}\widehat{F},\widehat{\xi },\widehat{\eta }\big) $-structure of index~$k$.
\end{theorem}

We conclude that the class of natural nonsymmetric metrics naturally
generalizes the concepts of (pseudo) Riemannian metrics and of almost
Hermitian/hyperbolic structures.

\subsection{Main theorems for d-connections metric compatibility}\label{sec5.2}

For a f\/ixed d-connection ${}_{\circ }\mathbf{D}=({}_{\circ }D_{k},{}_{\circ
}D_{c})=\big( {}_{\circ }L_{jk}^{i},{}_{\circ }C_{bc}^{a}\big)$,
we establish the existence and arbitariness of the d-connections which are
compatible with a nonsymmetric d-metric $\check{\mathbf{g}}=\mathbf{g}+
\mathbf{a}$. Our aim is to redef\/ine the A.~Kawaguchi metrization procedure~\cite{kaw1,kaw2,kaw3} considered in Finsler geometry and developed for
Lagrange spaces (including their noncommutative metric gene\-ra\-li\-za\-tions)~\cite{nsmat,mhat,ma1987} for nonholonomic manifolds enabled with nonsymmetric
metric structures~\cite{avnsm01,vrf01,vrf02}.

Let us consider a system of tensorial equations for unknown d-tensors ${}_{h}\mathbf{B} = \{B_{rk}^{s}\}$ and ${}_{v}\mathbf{B}= \{B_{ec}^{b}\},$
\begin{gather}
{}_{h}^{+}O{}_{h}\mathbf{B}={}_{h}\mathbf{U},\qquad {}_{v}^{+}O{}_{v}\mathbf{B}
={}_{v}\mathbf{U}, \qquad
{}_{h}^{+}\Psi {}_{h}\mathbf{B}={}_{h}\widetilde{\mathbf{U}},\qquad
{}_{v}^{+}\Psi {}_{v}\mathbf{B}={}_{v}\widetilde{\mathbf{U}},  \notag
\\
 {l}_{i}^{r}  \big( a_{sj}  B_{rk}^{s}+{}_{\circ
}D_{k}a_{rj}\big) =0,\qquad {l}_{a}^{b} \big( a_{bd}  B_{ec}^{b}+{}_{\circ }D_{c}a_{ed}\big) =0,  \notag \\
 {l}_{s}^{i}  \big( {m}_{j}^{r}  B_{rk}^{s}+{}_{\circ }D_{k}%
{l}_{j}^{s}\big) =0,\qquad {l}_{b}^{a}  \big( {m}%
_{d}^{e}  B_{ec}^{b}+{} _{\circ }D_{c}{l}_{d}^{b}\big) =0,  \label{dtenseq}
\end{gather}
where ${}_{h}\mathbf{U}=\{U_{rk}^{i}\}$, ${}_{v}\mathbf{U}=\{U_{ec}^{b}\}$ and $%
{}_{h}\widetilde{\mathbf{U}} = \{\widetilde{U}_{rk}^{s}\}$,
${}_{v} \widetilde{\mathbf{U}} = \{\widetilde{U}_{ec}^{b}\}$ are given by
formulas
\begin{equation}
2  U_{jk}^{i}=-g^{ir}{}_{\circ }D_{k}g_{rj},\qquad 2 U_{ec}^{b}=-g^{bd}{}_{\circ
}D_{c}g_{de},  \label{faux1}
\end{equation}%
and%
\begin{gather}
2\widetilde{U}_{jk}^{i}  = -\big( \check{a}^{ir}{}_{\circ }D_{k}a_{rj}+3%
{l}_{s}^{i}{}_{\circ }D_{k}{l}_{j}^{s}-{}_{\circ }D_{k}{%
l}_{j}^{i}\big) ,  \notag \\
2\widetilde{U}_{ec}^{b}  = -\big( \check{a}^{bd}{}_{\circ }D_{c}a_{de}+3 {l}_{d}^{b}{}_{\circ }D_{c}{l}_{e}^{d}-{}_{\circ }D_{c}{%
l}_{e}^{b}\big) . \label{faux2}
\end{gather}

One holds:

\begin{theorem}
\label{thau4}A d-connection $\mathbf{D}=\big( L_{jk}^{i},
C_{bc}^{a}\big) ={}_{\circ }\mathbf{D+B}$ is compatible with the
nonsymmetric d-metric $\check{\mathbf{g}}=\mathbf{g}+\mathbf{a}$ on
$\mathbf{V}^{n+n}$ if it is defined by a deformation (distorsion) d-tensor $%
\mathbf{B}=\big({}_{h}\mathbf{B},{}_{v}\mathbf{B}\big) $ which is a
solution of d-tensor equations~\eqref{dtenseq} for the values $\mathbf{U}=
\big({}_{h}\mathbf{U},{}_{v}\mathbf{U}\big) $ and $\widetilde{\mathbf{U}}%
=\big({}_{h}\widetilde{\mathbf{U}},{}_{v}\widetilde{\mathbf{U}}\big) $ constructed from the coefficients of the nonsymmetric
metric and fixed d-connection ${}_{\circ }\mathbf{D}$ following,
respectively, formulas~\eqref{faux1} and~\eqref{faux2}.
\end{theorem}

\begin{proof}
We sketch the proof for the h-components (considerations for v-components
being si\-mi\-lar). The metricity conditions (\ref{nscomp}) are equivalent to
\begin{equation*}
{}_{\circ }D_{k}g_{ij}+g_{sr}B_{jk}^{r}+g_{rj}B_{sk}^{r}=0\qquad \mbox{and}\qquad
{}_{\circ }D_{k}a_{ij}+a_{sr}B_{jk}^{r}+a_{rj}B_{sk}^{r}=0.
\end{equation*}
Contracting the f\/irst equation with $g^{si}$ we get the f\/irst equation from (\ref{dtenseq}). Contracting the second equation with $\check{a}^{ir}$ and $%
{l}_{i}^{r}$ and taking into account the the compatibility of $\mathbf{D}$ with $\check{\mathbf{g}}$ means ${l}_{s}^{i} D_{k}
{l}_{j}^{s}=0$, one obtains the f\/irsts equations from the third and
forth rows in (\ref{dtenseq}) and the equation ${}_{h}^{+}\Psi{}_{h}\mathbf{B}={} _{h}\widehat{\mathbf{U}},$ where $2\widehat{U}_{jk}^{i}=-\big( \check{a}^{ir}{}_{\circ }D_{k}a_{rj}+{l}_{s}^{i}{}
_{\circ }D_{k}{l}_{j}^{s}\big) $. Following Proposition~\ref{prsoleq}, the last equation is equivalent to the f\/irst equation in the
second row in (\ref{dtenseq}) with ${}_{h}\widetilde{\mathbf{U}}={}_{h}
\widehat{\mathbf{U}}+2{}_{h}\theta {}_{h}\widehat{\mathbf{U}}$, where the
values (\ref{faux1}) are obtained by re-grouping the coef\/f\/icients.
\end{proof}

One can be proposed a further simplif\/ication of such geometric models:

\begin{definition}
A natural nonsymmetric metric of index $k$ is of elliptic (hyperbolic) type
if $\nu =-c_{0}^{2}$ ($\nu =c_{0}^{2}$), where $c_{0}$ is a nonzero constant.
\end{definition}

This def\/inition is suggested by Theorems~\ref{thsf1}, \ref{thsf2} and \ref{thsf3} and Proposition~\ref{prmc}:

\begin{corollary}
\label{corel}The function $\nu (x,y)$ in equation \eqref{aux9} is a nonzero
constant, $\nu (x,y)=\mp c_{0}^{2}$, if the d-connection $\mathbf{D}$ is
compatible with the normal nonsymmetric metric $\check{\mathbf{g}}$.
\end{corollary}

\begin{proof}
One follows from equations~(\ref{nscomp}) and Proposition~\ref{prmc} that
$(\mathbf{D}_{\alpha }\nu )a_{ij}=0$ which means that $(\mathbf{D}_{\alpha
}\nu )=0$ because $a_{ij}\check{a}^{ji}=n-k\neq 0.$ This is possible if $\nu
$ is a nonzero constant.
\end{proof}

One holds true the inverse statement of Theorem~\ref{thau4}:

\begin{theorem}
\label{thau5}For a natural nonsymmetric metric $\check{\mathbf{g}}$ of index
$k$, the values $\mathbf{U}=\big({}_{h}\mathbf{U},{}_{v}\mathbf{U}\big) $ and $\widetilde{\mathbf{U}} = \big({}_{h}\widetilde{\mathbf{U}} ,{}_{v}\widetilde{\mathbf{U}}\big) $ defined
respectively by formulas \eqref{faux1} and \eqref{faux2} satisfy the
equations
\begin{eqnarray}
{}_{h}^{+}O{}_{h}\mathbf{U}  = 0,\qquad {}_{h}^{+}\Psi {}_{h}\widetilde{\mathbf{U}}=0,\qquad
 {}_{h}^{-}\Psi {}_{h}\mathbf{U}={}_{h}^{-}O{}_{h} \widetilde{\mathbf{U}},  \notag \\
{}_{v}^{+}O{}_{v}\mathbf{U}  = 0,\qquad {}_{v}^{+}\Psi {} _{v}\widetilde{\mathbf{U}}=0,\qquad
 {}_{v}^{-}\Psi{}_{v}\mathbf{U}={}_{v}^{-}O{}_{v} \widetilde{\mathbf{U}}.  \label{aux10}
\end{eqnarray}
\end{theorem}

A proof of this Theorem consists from straightforward verif\/ications that~$\mathbf{U}$ and $\widetilde{\mathbf{U}}$ really solves the equations~(\ref{aux10}) for given values of Obata operators $\mathbf{O}$ and $\Psi $
constructed respectively from the components of symmetric and
skew-symmetric parts of nonsymmetric metric (in~\cite{ma1987} it is
contained a similar proof for the so-called nonsymmetric
Eisenhart--Lagrange metric on tangent bundles).

\begin{example}
\label{ex1}By direct computations, we can check that for any given
d-connection ${}_{\circ }\Gamma _{\ \beta \gamma }^{\alpha }=\big({}
_{\circ }L_{jk}^{i},{}_{\circ }C_{bc}^{a}\big) $ and nonsymmetric
d-metric $\check{\mathbf{g}}=\mathbf{g}+\mathbf{a}$ on $\mathbf{V}^{n+n}$
the d-connection ${}_{\ast }\Gamma _{\ \beta \gamma }^{\alpha }{=}\big({}_{\ast }L_{jk}^{i},{}_{\ast }C_{bc}^{a}\big) $, where
\begin{gather}
{}_{\ast }L_{jk}^{i} ={} _{\circ }L_{jk}^{i}+\frac{1}{2}[g^{ir}{} _{\circ
}D_{k}g_{rj}+  {}^{\pm }O_{sj}^{ir}(\check{a}^{st}{}_{\circ }D_{k}a_{tr}+3{l}%
_{t}^{s}{}_{\circ }D_{k}{l}_{r}^{t}-{}_{\circ }D_{k}{l}%
_{r}^{s})] , \notag \\
{}_{\ast }C_{bc}^{a} ={} _{\circ }C_{bc}^{a}+\frac{1}{2}[g^{ah}{}_{\circ
}D_{c}g_{hb}+  {}^{\pm }O_{eb}^{ah}(\check{a}^{ed}{}_{\circ }D_{c}a_{dh}+3{l}%
_{d}^{e}{}_{\circ }D_{c}{l}_{h}^{d}-{}_{\circ }D_{c}{l}%
_{h}^{e})]  \label{aux11}
\end{gather}%
is d-metric compatible, i.e.\ satisf\/ies the conditions ${}_{\ast }\mathbf{D}
\check{\mathbf{g}}=0$.
\end{example}

In a more general case, one holds:

\begin{theorem}
\label{thmain}The set of d-connections $\mathbf{D}={}_{\circ }\mathbf{D}+\mathbf{B}$
being generated by deformations of an arbitrary fixed d-connection ${}_{\circ }\mathbf{D}$ in order to be compatible with a given nonsymmetric
d-metric $\check{\mathbf{g}}=\mathbf{g}+\mathbf{a}$ on $\mathbf{V}^{n+n}$
is defined by distorsion d-tensors $\mathbf{B}=\big({}_{h}\mathbf{B},{}_{v}\mathbf{B}\big) $ of type
\begin{equation}
{}_{h}\mathbf{B}={}_{h}\mathbf{U}+{}_{h}^{+}O\big({}_{h}\widetilde{\mathbf{U}}+{}_{h}^{+}\Psi   \mathbf{Y}\big) \qquad \mbox{and}\qquad {}_{v}\mathbf{B}={}_{v}%
\mathbf{U}+{}_{v}^{+}O\big({}_{v}\widetilde{\mathbf{U}}+{}_{v}^{+}\Psi
\mathbf{Z}\big) ,  \label{aux12a}
\end{equation}%
or
\begin{equation}
{}_{h}\mathbf{B}={}_{h}\widetilde{\mathbf{U}} +{} _{h}^{+}\Psi \big(
{}_{h}\mathbf{U}+{}_{h}^{+}O  \mathbf{Y}\big) \qquad \mbox{and}\qquad {} _{v}\mathbf{B}=
{}_{v}\widetilde{\mathbf{U}} +{}_{v}^{+}\Psi \big({}_{v}\mathbf{U} +{}_{v}^{+}O \mathbf{Z}\big) ,  \label{aux12b}
\end{equation}%
where $\mathbf{Y}$ and $\mathbf{X}$ are arbitrary d-tensor fields of type $(1,2)$.
\end{theorem}

\begin{proof}
It follows from Theorems~\ref{thau4} and~\ref{thau5}.
\end{proof}

For the statements of Example \ref{ex1}, we have:

\begin{remark}
The formulas for the metric compatible d-connection ${}_{\ast }\mathbf{\Gamma }_{\ \beta \gamma }^{\alpha }$ (\ref{aux11}) consist a particular
case when the deformation d-tensor $\mathbf{B}=\big({}_{h}\mathbf{B},{}_{v}\mathbf{B}\big) $ is computed by (\ref{aux12a}), or (\ref{aux12b}),
for $\mathbf{Y}=0$ and $\mathbf{Z}=0$.
\end{remark}

From the Corollary~\ref{corel} and Theorem~\ref{thmain}, we get:

\begin{conclusion}
\label{conclthmain}The set of all d-connections $\mathbf{\Gamma }_{\
\beta \gamma }^{\alpha } =\big( L_{jk}^{i},C_{bc}^{a}\big) $ being
compatible to a nonsymmetric metric $\check{\mathbf{g}}$ of elliptic/hyperbolic type on $\mathbf{V}^{n+n}$ is parametrized by formulas
\begin{equation}
L_{jk}^{i}={}_{\ast }L_{jk}^{i}+{}^{+}O_{sj}^{ir}{}^{+}\Psi _{pr}^{sm}
\mathbf{Y}_{mk}^{p}\qquad \mbox{and}\qquad C_{bc}^{a}={}_{\ast }C_{bc}^{a}+{}^{+}O_{eb}^{ah}{}^{+}\Psi _{fh}^{ed}  \mathbf{Z}_{dc}^{f},  \label{aux}
\end{equation}
where ${}_{\ast }L_{jk}^{i}$ and ${}_{\ast }C_{bc}^{a}$ are given
respectively by formulas \eqref{aux11}, ${}_{h}^{+}\Psi =\{{}^{+}\Psi
_{pr}^{sm}\}$ and ${}_{v}^{+}\Psi =\{{}^{+}\Psi _{fh}^{ed}\}$ and $\mathbf{Y}=
\{Y_{mk}^{p}\}$ and $\mathbf{Z}=\{Z_{dc}^{f}\}$ are correspondingly
arbitrary horizontal and vertical d-tensors of type $(1,2)$.
\end{conclusion}

To chose a parametrization that $\nu$ is constant is the simplest way to
prove formulas~(\ref{aux}) def\/ining the set of all d-connections being
compatible to a given nonsymmetric metric.

Finally, we note that because ${}_{\circ }\Gamma _{\ \beta \gamma }^{\alpha
}=\big({}_{\circ }L_{jk}^{i},{}_{\circ }C_{bc}^{a}\big) $ is an
arbitrary d-connection, we can chose it to be an important one for certain
physical or geometrical problems. For instance, in Section~\ref{snsfls} we
shall consider that ${}_{\circ }\Gamma _{\ \beta \gamma }^{\alpha }$ is
def\/ined by the Cartan d-connection constructed for Lagrange/Finsler spaces
for a symmetric d-metric $g$ corresponding to the Lagrange/Finsler
d-metric which will allow us to perform certain nonsymmetric
generalizations of such geometries. In our further researches, we are going
to consider certain exact solutions in gravity with nonholonomic variables
def\/ining a corresponding ${}_{\circ }\Gamma _{\ \beta \gamma }^{\alpha }$ and
then deformed to nonsymmetric conf\/igurations following formulas (\ref{aux11}) and/or (\ref{aux}).

\section{Nonsymmetric gravity and nonholonomic frames}\label{sec6}

A nonsymmetric gravitational theory (NGT) based on decompositions of the
nonsymmetric metric $\check{g}_{\mu \nu }$ and af\/f\/ine connection $\Gamma
_{\beta \gamma }^{\alpha }$ was elaborated in series of work by J. Mof\/fat
and co--authors, see \cite{moff1,moff1a,moffrev,nsgtjmp,moffncqg,moff0505326}
and references therein\footnote{we note that in this paper we use a dif\/ferent system of denotations; the
f\/inal version of NGT proposed by J.~Mof\/fat's group is free of ghosts,
tachions and higher-order poles in the propagator in the linear
approximation on Minkowski space; an expansion of the general nonsymmetric
metric about an arbitrary Einstein background metric yields f\/ield equations
to f\/irst order in the skew-symmetric part of metric, which are free of
coupling to un-physical (negative energy) modes; the solutions of such
gravitational f\/ield equations have consistent asymptotic boundary
conditions; here, it should be noted that in the mentioned works a set of
theoretical and experimental data were explained and in consistent way by
NGT and its further modif\/ications.}.

In this section, we show how NGT can be formulated on N-anholonomic
manifolds where an additional geometric structure (the N-connection) is
present and the geometric constructions can be equivalently (at least at
classical level) performed in N-adapted or not N-adapted forms.

We note that for a class of geometric and physical models the
N-connecti\-on coef\/f\/icients are induced by certain subsets of generic
of\/f-diagonal coef\/f\/icients of the metric. In this approach, the
N-connection splitting results in a nonholonomic decompositon of geometric
objects with respect to certain frames with mixed holonomic-nonholonomic
basic vectors which may be convenient for constructing, for instance, new
classes of exact solutions, to def\/ine spacetimes with generalized
symmetries, nonholomogeneity and ef\/fective local anisotropy, or in order to
elaborate certain models of gauge type and/or geometric/deformation
quantization.

In a more general context, we can consider that the N-connection is def\/ined
by an additional geometric structure (independent from the metric and linear
connection structures), like in gene\-ra\-lized Lagrange--Finsler theories, when
certain geometric or physical/mechanical theories are def\/ined on
nonholonomic manifolds. Nevertheless, even in such cases, following the
method of metrization of d-connections considered in the previous section,
we can redef\/ine the constructions for nonholonomic (pseudo) Riemann or
Riemann--Cartan with additional ef\/fective f\/ield interactions and
nonholonomic constraints. As a matter of principle, we can use the Levi-Civita connection working with non N-adapted geometric objects.

\subsection{On (not) N-adapted models of NGT}\label{sec6.1}

Let us consider a d-connection $\mathbf{\Gamma }_{\ \beta \gamma
}^{\alpha }$ not obligatory compatible to a nonsymmetric metric $\check{\mathbf{g}}=\{\check{\mathbf{g}}_{\alpha \beta }\}$ (\ref{hvm}) on
$\mathbf{V}^{2+2}$ enabled with an arbitrary N-connection structure $%
\mathbf{N}=\{N_{i}^{a}\}$ (\ref{coeffnc})\footnote{For our purposes, we consider four dimensional spacetimes with 2+2
nonholonomic splitting.}. We can introduce nonsymmetric gravitational
equations following the approach elaborated in~\cite{moff1a}, but in
our case working with respect to N-adapted bases (\ref{dder}) and (\ref{ddif}).

For any system of reference, we can write
\begin{equation*}
g_{\mu \nu }=\check{g}_{(\mu \nu )}=\frac{1}{2}\left( \check{g}_{\mu \nu }+%
\check{g}_{\mu \nu }\right) ,\qquad a_{\mu \nu }=\check{g}_{[\mu \nu ]}=\frac{1}{2%
}\left( \check{g}_{\mu \nu }-\check{g}_{\mu \nu }\right) ,\qquad \Gamma _{\mu \nu
}^{\lambda }=\Gamma _{(\mu \nu )}^{\lambda }+\Gamma _{\lbrack \mu \nu
]}^{\lambda }.
\end{equation*}
Introducing the unconstrained (nonsymmetric) af\/f\/ine connection%
\begin{equation}
W_{\mu \nu }^{\lambda }\doteqdot \Gamma _{\mu \nu }^{\lambda }-\frac{2}{3}%
\delta _{\mu }^{\lambda }W_{\nu },  \label{moffatc}
\end{equation}%
where
 $W_{\nu }=\frac{1}{2}\big( W_{\mu \lambda }^{\lambda }
 -W_{\lambda \mu}^{\lambda }\big)$,  %
which leads to the condition $\Gamma _{\mu }=\Gamma _{\lbrack \mu \lambda
]}^{\lambda }=0.$ We note that the coef\/f\/icients of $\Gamma _{\mu \nu
}^{\lambda }$ and $W_{\mu \nu }^{\lambda }$ are given with respect to
N-adapted bases (\ref{dder}) and (\ref{ddif}) but we do not use for such
values ``boldfaced'' symbols because we have not supposed that these linear
connections are adapted to the N-connection splitting (\ref{whitney}).

The contracted curvature tensors for the above linear connections are
related by formulas
\begin{equation*}
{}_{W}R_{\mu \nu }={}_{\Gamma }R_{\mu \nu }+\frac{2}{3}\mathbf{e}_{[\nu
}W_{\mu ]},
\end{equation*}%
where
\begin{equation*}
{}_{\Gamma }R_{\mu \nu }=\mathbf{e}_{\beta }\Gamma _{\mu \nu }^{\beta }-%
\frac{1}{2}\big( \mathbf{e}_{\nu }\Gamma _{(\mu \lambda )}^{\lambda }+%
\mathbf{e}_{\mu }\Gamma _{(\nu \lambda )}^{\lambda }\big) -\Gamma _{\alpha
\nu }^{\beta }\Gamma _{\mu \beta }^{\alpha }+\Gamma _{(\alpha \lambda
)}^{\lambda }\Gamma _{\mu \nu }^{\alpha }.
\end{equation*}

The f\/ield equations of N-anholonomic NGT in presence of a source $\Upsilon
_{\mu \nu }$ for matter f\/ields are%
\begin{gather}
{}_{W}G_{\mu \nu }+\lambda \check{\mathbf{g}}_{\mu \nu }+\frac{\mu ^{2}}{4}
\mathbf{S}_{\mu \nu }-\frac{1}{6}{}_{P}G_{\mu \nu } =8\pi \Upsilon _{\mu
\nu },   \notag \\
2 \mathbf{e}_{\nu }\big( \sqrt{|\check{\mathbf{g}}|}\check{\mathbf{g}}%
^{[\nu \mu ]}\big) =\sqrt{|\check{\mathbf{g}}|}\check{\mathbf{g}}^{(\nu
\mu )}W_{\nu },  \label{nsgfe}\\
\big( \sqrt{|\check{\mathbf{g}}|}\big) ^{-1}\mathbf{e}_{\sigma }\big(
\sqrt{|\check{\mathbf{g}}|}\check{\mathbf{g}}^{\mu \nu }\big) +\check{\mathbf{g}}^{\rho \nu }W_{\rho \sigma }^{\mu }-\check{\mathbf{g}}^{\mu \nu
}W_{\rho \sigma }^{\rho }+ 
\frac{W_{\beta }}{6}\big( \delta _{\sigma }^{\nu }\check{\mathbf{g}}^{(\mu
\beta )}+\delta _{\sigma }^{\mu }\check{\mathbf{g}}^{(\nu \beta )}\big) +
\frac{2}{3}\delta _{\sigma }^{\nu }\check{\mathbf{g}}^{\mu \rho }W_{[\rho
\beta ]}^{\beta } =0,  \notag
\end{gather}%
where $|\check{\mathbf{g}}|\doteqdot \det |\check{\mathbf{g}}_{\mu \nu }|$,
$\check{\mathbf{g}}^{\mu \nu }\check{\mathbf{g}}_{\sigma \nu }=\check{\mathbf{g}}^{\nu \mu }\check{\mathbf{g}}_{\nu \sigma }=\delta _{\sigma }^{\mu }$ (we
use boldface indices for the nonsymmetric metric and bases (\ref{dder}) and (\ref{ddif}) because they can be adapted to the N-connection structure even
a general linear connection and related tensors are not distinguished)$;\
\lambda $ is the cosmological constant and $\mu ^{2}$ is an additional
cosmological constant associated to to $\mathbf{a}_{\mu \nu }$ (there are
used the physical units when the gravitational and vacuum speed are stated
to be dimensionless and equal to unity);
\begin{gather}
\mathbf{S}_{\mu \nu }  \doteqdot  \mathbf{a}_{\mu \nu }+\check{\mathbf{g}}%
^{[\rho \sigma ]}\left( \check{\mathbf{g}}_{\mu \sigma }\check{\mathbf{g}}%
_{\rho \nu }+\frac{1}{2}\mathbf{a}_{\sigma \rho }\check{\mathbf{g}}_{\mu \nu
}\right),  \notag\\
{}_{W}G_{\mu \nu } \doteqdot {} _{W}R_{\mu \nu }-\frac{1}{2}\check{\mathbf{g}}_{\mu \nu }{}_{W}\overleftarrow{R},\qquad {}_{P}G_{\mu \nu }\doteqdot
P_{\mu \nu }-\frac{1}{2}\check{\mathbf{g}}_{\mu \nu } P,\label{aux12}
\end{gather}%
for $P_{\mu \nu }\doteqdot W_{\mu }W_{\nu }$ and $P\doteqdot \check{\mathbf{g}}^{\mu \nu }P_{\mu \nu }=\check{\mathbf{g}}^{(\mu \nu )}P_{\mu \nu },$ when
the scalar curvatures are def\/ined respec\-tively ${}_{W}\overleftarrow{R}%
\doteqdot \check{\mathbf{g}}^{\mu \nu }{}_{W}R_{\mu \nu }$, ${}_{\Gamma }%
\overleftarrow{R}\doteqdot \check{\mathbf{g}}^{\mu \nu }{}_{\Gamma }R_{\mu
\nu },\dots$ We note that, in general, the h- and v-components of~$\check{\mathbf{g}}^{[\rho \sigma ]}$ are dif\/ferent from $\widetilde{a}^{-1}$ (\ref{aux00}) because we have not yet introduced here metric d-connec\-tions and
complete N-adapted and d-tensor calculus.

The matter f\/ields d-tensor $\Upsilon _{\mu \nu }$ from the f\/irst equation
in (\ref{nsgfe}) is constrained to satisfy the so-called matter response
equations,
\begin{equation}
\check{\mathbf{g}}_{\mu \rho }\mathbf{e}_{\nu }(\sqrt{|\check{\mathbf{g}}|}%
\Upsilon ^{\mu \nu })+\check{\mathbf{g}}_{\rho \mu }\mathbf{e}_{\nu }(\sqrt{|%
\check{\mathbf{g}}|}\Upsilon ^{\nu \mu })+\left( \mathbf{e}_{\nu }\check{\mathbf{g}}_{\mu \rho }+\mathbf{e}_{\mu }\check{\mathbf{g}}_{\rho \nu }-%
\mathbf{e}_{\rho }\check{\mathbf{g}}_{\mu \nu }\right) \sqrt{|\check{\mathbf{g}}|}\Upsilon ^{\nu \mu }=0,  \label{conslaw}
\end{equation}%
which is a consequence of the generalized Bianchi identities%
\begin{equation*}
\mathbf{e}_{\mu }\big[ \sqrt{|\check{\mathbf{g}}|}\check{\mathbf{g}}^{\mu
\nu }{}_{\Gamma }G_{\rho v}+\sqrt{|\check{\mathbf{g}}|}\check{\mathbf{g}}%
^{\nu \mu }{}_{\Gamma }G_{v\rho }\big] +\sqrt{|\check{\mathbf{g}}|}{}_{\Gamma }G_{\mu v}\mathbf{e}_{\rho }\check{\mathbf{g}}^{\mu \nu }=0,
\end{equation*}%
see a detailed study in~\cite{legmoff,moff1a}, where a variational
proof of equations~(\ref{nsgfe}) and~(\ref{conslaw}) is formulated by f\/ixing
a corresponding coef\/f\/icient before $P_{\mu \nu }$ in order to yield a
consistent theory with ghost and tachyon free perturbative solutions to the
f\/ield equations.

It should be emphasized that dif\/ferent models of nonsymmteric theory of
gravity were elaborated in such a way that the induced general linear
connection is not metric compatible. In our approach, we shall prove that it
is possible to elaborate the nonsymmetric gravity theory in a general metric
compatible form by using corresponding classes of d-connections. This is
very important from physical point of view (there is not a well accepted
interpretation of nonmetricity f\/ields) and presents certain interests from
the viewpoint of the Ricci f\/low theory with nonholonomic constraints \cite%
{vrf02,avnsm01} when in the simplest approach a symmetric metric can evolve
into a nonsymetric one, and inversely, but preserving the general metric
compatibility of li\-near connection which is very important for def\/inition of
conservation laws and related physical values.

\begin{theorem}
\label{theormcngt}We obtain a canonical d-metric compatible and N-adapted
nonholonomic NGT completely defined by a N-connection $\mathbf{N}%
=\{N_{i}^{a}\}$ and d-metric $\check{\mathbf{g}}=\mathbf{g}+\mathbf{a}$ \eqref{hvm} of
elliptic/hyperbolic type if we chose instead of arbitrary affine
connection $\Gamma _{\mu \nu }^{\lambda }$ the metric compatible
d-connection ${}_{\ast }\Gamma _{\ \beta \gamma }^{\alpha }=\big({}_{\ast
}L_{jk}^{i},{}_{\ast }C_{bc}^{a}\big) $ \eqref{aux11}.
\end{theorem}

\begin{proof}
We sketch the idea of the proof which can be obtained by a N-adapted
variational calculus, similar to that from \cite{legmoff} when instead of
partial derivatives there are used the ``N-elongated'' partial derivatives $%
\mathbf{e}_{\rho }$ (\ref{dder}), varying independently the d-f\/ields $\check{\mathbf{g}}=\mathbf{g}+\mathbf{a}$ and ${}_{\ast }\mathbf{\Gamma }_{\ \beta \gamma
}^{\alpha }.$ In this case, $\check{a}=( \check{a}^{ij}) $ does
not depend on the choice of f\/ields $\widehat{\xi }$, see (\ref{aux00}), and
we can write $\check{\mathbf{g}}^{[\rho \sigma ]}=\check{\mathbf{a}}^{\rho
\sigma }=[\check{a}^{ij},\check{a}^{cb}],$ where $\check{a}^{ij}=-\check{a}%
^{ji}$ and $\check{a}^{cb}=-\check{a}^{bc}.$ Instead of an af\/f\/ine connection
(\ref{moffatc}) we work with metric d-connections,
\begin{equation}
{}_{\ast }\mathbf{W}_{\mu \nu }^{\lambda }\doteqdot {}_{\ast }\mathbf{\Gamma
}_{\mu \nu }^{\lambda }-\frac{2}{3}\delta _{\mu }^{\lambda }{}_{\ast }%
\mathbf{W}_{\nu },  \label{stbfc}
\end{equation}%
where ${}_{\ast }\mathbf{W}_{\nu }=\frac{1}{2}\big({} _{\ast }\mathbf{W}%
_{\mu \lambda }^{\lambda }-{}_{\ast }\mathbf{W}_{\lambda \mu }^{\lambda
}\big),$ and we use boldface symbols. The corresponding Ricci d-tensors
are related by formulas
\begin{equation*}
{}_{W}^{\ast }\mathbf{R}_{\mu \nu }={}_{\Gamma }^{\ast }\mathbf{R}_{\mu \nu
}+\frac{2}{3}\mathbf{e}_{[\nu }{}^{\ast }\mathbf{W}_{\mu ]},
\end{equation*}
where
\begin{equation*}
{}_{\Gamma }^{\ast }\mathbf{R}_{\mu \nu }=\mathbf{e}_{\beta }{}_{\ast
}\Gamma _{\mu \nu }^{\beta }-\frac{1}{2}\big( \mathbf{e}_{\nu }{}_{\ast
}\Gamma _{(\mu \lambda )}^{\lambda }+\mathbf{e}_{\mu }{}_{\ast }\Gamma
_{(\nu \lambda )}^{\lambda }\big) -{}_{\ast }\Gamma _{\alpha \nu }^{\beta
}{}_{\ast }\Gamma _{\mu \beta }^{\alpha }+{}_{\ast }\Gamma _{(\alpha \lambda
)}^{\lambda }{}_{\ast }\Gamma _{\mu \nu }^{\alpha }.
\end{equation*}%
has h- and v-components of type (\ref{driccic}).

The canonical N-adapted f\/ield equations for the nonholonomic NGT are
\begin{gather}
{}_{W}^{\ast }\mathbf{G}_{\mu \nu }+\lambda \check{\mathbf{g}}_{\mu \nu }+%
\frac{\mu ^{2}}{4}{}_{\ast }\mathbf{S}_{\mu \nu }-\frac{1}{6}{}_{P}^{\ast }%
\mathbf{G}_{\mu \nu }  = 8\pi \mathbf{\Upsilon }_{\mu \nu }, \notag
\\
2 \mathbf{e}_{\nu }\big( \sqrt{|\check{\mathbf{g}}|}\check{\mathbf{a}}%
^{[\nu \mu ]}\big)  = \sqrt{|\check{\mathbf{g}}|}\check{\mathbf{g}}^{(\nu
\mu )}{}_{\ast }\mathbf{W}_{\nu },  \notag \\
\big( \sqrt{|\check{\mathbf{g}}|}\big) ^{-1}\mathbf{e}_{\sigma }\big(
\sqrt{|\check{\mathbf{g}}|}\check{\mathbf{g}}^{\mu \nu }\big) +\check{\mathbf{g}}^{\rho \nu }{}_{\ast }\mathbf{W}_{\rho \sigma }^{\mu }-\check{\mathbf{g}}^{\mu \nu }{}_{\ast }\mathbf{W}_{\rho \sigma }^{\rho }   \notag
\\
\qquad{}+\frac{{}_{\ast }\mathbf{W}_{\beta }}{6}\big( \delta _{\sigma }^{\nu }%
\check{\mathbf{g}}^{(\mu \beta )}+\delta _{\sigma }^{\mu }\check{\mathbf{g}}%
^{(\nu \beta )}\big) +\frac{2}{3}\delta _{\sigma }^{\nu }\check{\mathbf{g}}%
^{\mu \rho }{}_{\ast }\mathbf{W}_{[\rho \beta ]}^{\beta }  = 0,  \label{nscgfe}
\end{gather}%
where the formulas for geometric objects and conservation laws are def\/ined,
respectively, similarly to (\ref{aux12}) and (\ref{conslaw}) but for
boldfaced d-connections (\ref{stbfc}). These equations can be derived from
the Lagrangian density
\begin{gather*}
{} _{\ast }\mathcal{L}_{\rm NGT} ={} _{\ast }\mathcal{L}_{R}+\mathcal{L}_{M}, \\
{}_{\ast }\mathcal{L}_{R} =\sqrt{|\check{\mathbf{g}}|}\check{\mathbf{g}}%
^{\rho \nu }{}_{W}^{\ast }\mathbf{R}_{\mu \nu }-2\lambda \sqrt{|\check{\mathbf{g}}|}-\frac{\mu ^{2}}{4}\sqrt{|\check{\mathbf{g}}|}\check{\mathbf{g}}%
^{\mu \nu }\mathbf{a}_{\mu \nu }-\frac{1}{6}\check{\mathbf{g}}^{\mu \nu }{}_{\ast }\mathbf{W}_{\mu }{}_{\ast }\mathbf{W}_{\nu }, \\
\mathcal{L}_{M} =-8\pi \check{\mathbf{g}}^{\mu \nu }\mathbf{\Upsilon }%
_{\mu \nu }
\end{gather*}%
following a N-adapted variational calculus.
\end{proof}

The above considerations motivate the concepts:

\begin{definition}
A Mof\/fat gravity (spacetime) model is def\/ined by a nonsymmetric metric $%
\check{\mathbf{g}}=\{\check{\mathbf{g}}_{\alpha \beta }\}$ (\ref{hvm}) and
an af\/f\/ine connection $W_{\mu \nu }^{\lambda }\doteqdot \Gamma _{\mu \nu
}^{\lambda }-\frac{2}{3}\delta _{\mu }^{\lambda }W_{\nu }$~(\ref{moffatc})
solving the NGT f\/ield equations~(\ref{nsgfe}).
\end{definition}

In a more general case, we have

\begin{definition}
A canonical nonholonomic Eisenhart--Mof\/fat gravity (spacetime) model is
def\/ined by a f\/ixed N-connection structure $\mathbf{N}$ and a nonsymmetric
metric $\check{\mathbf{g}}=\{\check{\mathbf{g}}_{\alpha \beta } \}$ (\ref{hvm}%
) and a~metric compatible d-connection ${}_{\ast }\Gamma _{\ \beta \gamma
}^{\alpha }=\big({}_{\ast }L_{jk}^{i},{}_{\ast }C_{bc}^{a}\big) $ (\ref{aux11}) solving the N-adapted NGT f\/ield equations (\ref{nscgfe}).
\end{definition}

As a matter of principle, following the Kawaguchi's metrization method we
can work with various classes of metric or nonmetric d-connections:

\begin{remark}
Analogous of Theorem~\ref{theormcngt} can be formulated and proven for any
metric noncompatible d-connection ${}_{\circ }\mathbf{D}=({}_{\circ
}D_{k},{}_{\circ }D_{c})=\big({}_{\circ }L_{jk}^{i},{}_{\circ
}C_{bc}^{a}\big) $, or for any metric compatible d-connection $\mathbf{\Gamma }_{\ \beta \gamma }^{\alpha }=\big( L_{jk}^{i},C_{bc}^{a}\big) $
(\ref{aux}). In all cases, we get N-adapted models of NGT but for ${}_{\circ }\mathbf{D}$ we generate geometric constructions for nonmetric
spaces and for $\mathbf{\Gamma }_{\ \beta \gamma }^{\alpha }$ the models
depend on d-tensor f\/ields of type $(1,2)$ for which one has to provide
additional geometric and physical motivations.
\end{remark}

Here we emphasize that it is preferred to work with metric compatible
connection for global def\/inition of spinors and noncommutative versions with
Dirac operators in dif\/ferent models of gravity because metric compatibility
results in compatible structure between spinor connections and generating
Clif\/ford structures bases.

From the last Theorem and Remark, one follows:

\begin{corollary}
If the N-connection structure is induced, for instance, by the
off-diagonal coef\-ficients of the symmetric part of the nonsymmetric metric,
and the canonical nonholonomic Eisenhart--Moffat gravity (which is a metric
compatible theory) is equivalent to the Moffat's gravity theory (which was
performed in a metric noncompatible form).
\end{corollary}

\begin{proof}
We state that the symmetric part $\mathbf{g}=\mathbf{g}_{\alpha \beta }%
\mathbf{e}^{\alpha }\otimes \mathbf{e}^{\beta }=g_{ij}e^{i}\otimes
e^{j}+g_{ab}\mathbf{e}^{a}\otimes \mathbf{e}^{b}$ in (\ref{hvm}) with
respect to a coordinate base $e^{\alpha }=du^{\alpha }=(dx^{i},dy^{a}),$ is
given in the form
\begin{equation*}
\mathbf{g}=\underline{g}_{\alpha \beta }\left( u\right) du^{\alpha }\otimes
du^{\beta }  
\end{equation*}%
where%
\begin{equation*}
\underline{g}_{\alpha \beta }=\left[
\begin{array}{cc}
g_{ij}+N_{i}^{a}N_{j}^{b}h_{ab} & N_{j}^{e}h_{ae} \\
N_{i}^{e}h_{be} & h_{ab}%
\end{array}%
\right] .  
\end{equation*}%
induces the coef\/f\/icients of N-connection $\mathbf{N}=N_{i}^{a}(u)dx^{i}%
\otimes \partial /\partial y^{a}$ (\ref{coeffnc}). The next step, is to take
${}_{\circ }\Gamma _{\ \beta \gamma }^{\alpha }={}_{n}\mathbf{\Gamma }_{\
\beta \gamma }^{\alpha },$ see Def\/inition~\ref{defncdc}, in ${}_{\ast }\Gamma_{\ \beta \gamma }^{\alpha }.$\footnote{In a more general, or special, approach, we can use the d-connections (\ref{candcon}), or (\ref{ccd}).} Using the deformation of connection ${}_{n}\mathbf{\Gamma }_{\ \beta \gamma }^{\alpha }= \Gamma _{\ \beta \gamma
}^{\alpha }+S_{\ \beta \gamma }^{\alpha }$, we can redef\/ine the N-adapted
f\/ield equations (\ref{nscgfe}) in terms of connection $\Gamma _{\ \beta
\gamma }^{\alpha }$ and~$S_{\ \beta \gamma }^{\alpha }.$ The last term can
be encoded in terms of d-tensor f\/ields of type $(1,2)$. This way a model of
metric compatible and N-adapted NGT is transformed in a Mof\/fat type model
of gravity.
\end{proof}

The NGT was proven to generate physically consistent models using linear
approximations for the nonsymmetric metric about Minkowski and Einstein
spaces, see \cite{legmoff,moff1a} and references therein. Various types of
approximations can be performed following a corresponding N-adapted
calculus with respect to a nonholonomic background. Depending on the type of
background and constraints on nonsymmetric metric components, we obtain
dif\/ferent ef\/fective models of scalar/vector/tensor gravity, with
variable/running physical constants which are intensively exploited in
modern cosmology, for instance, see \cite{moff0505326,prok}. To work with
nonholonomic backgrounds presents a substantial interest both from
conceptual and technical point of views: we can `extract' from NGT new
classes of nonholonomic Einstein, generalized Finsler--Lagrange, \dots spaces~\cite{vrfg}, establish certain new links and develop new methods in
geometric quantization~\cite{vqgr3} and noncommutative gravity~\cite{moffncqg,vggr,vncg}, as well to elaborate new methods of constructing exact
solutions in Einstein and generalized gravity theories and Ricci f\/lows \cite{vrfg,vrfsol1,vvisrf1,vvisrf2,vrf04,vrf05}.

\subsection[Expansion of field equations in NGT with respect to nonholonomic
backgrounds]{Expansion of f\/ield equations in NGT with respect\\ to nonholonomic
backgrounds}\label{sec6.2}

Let us consider the expansion of the f\/ield equations (\ref{nscgfe}) for $%
\check{\mathbf{g}}=\mathbf{g}+\mathbf{a}$ around a background spacetime def\/ined by a symmetric
d-connection $\mathbf{g}=\{\mathbf{g}_{\alpha \beta }\}$ and a metric
compatible d-connection~$\widehat{\mathbf{\Gamma }}_{\ \beta \gamma
}^{\alpha }$ def\/ined by $\mathbf{N}$ and $\mathbf{g}$ (it can be a normal,
the canonical d-connection, the Cartan or another one) and denote $\mathbf{%
\check{g}}_{[\alpha \beta ]}=\mathbf{a}_{\alpha \beta }.$\footnote{In order to elaborate a consistent ``perturbation'' theory on f\/ields $\mathbf{a%
}_{\alpha \beta },$ we may consider that such nonsymmetric deformations of
metrics are def\/ined by an additional, or the same Newton constant (but under
nonholonomic Ricci f\/lows) used as a small parameter. Such decompositions are
used for def\/inition of ``week'' gravitational waves in general relativity and
in ``perturbative'' models of quasi-classical gravity; in a more general
approach, for nonsymmmetric metrics, but in not N-adapted forms, such
constructions were introduced and developed in a series of works by J.~Mof\/fat and co-authors \cite{moff1,moff1a,moffrev,nsgtjmp}.} We shall compute
\begin{gather*}
\check{\mathbf{g}}_{\alpha \beta } =\mathbf{g}_{\alpha \beta }+{}^{1}%
\mathbf{g}_{\alpha \beta }+\cdots,\qquad \mathbf{a}_{\alpha \beta }={}^{1}\mathbf{a}%
_{\alpha \beta }+{}^{2}\mathbf{a}_{\alpha \beta }+\cdots, \\
\mathbf{\Gamma }_{\ \beta \gamma }^{\alpha } =\widehat{\mathbf{\Gamma }}%
_{\ \beta \gamma }^{\alpha }+{}^{1}\mathbf{\Gamma }_{\ \beta \gamma
}^{\alpha }+\cdots,\qquad \mathbf{W}_{\mu }={}^{1}\mathbf{W}_{\mu }+{}^{2}\mathbf{W%
}_{\mu }+\cdots.
\end{gather*}%
Substituting into f\/ield equations, for $\lambda =0$ and $\mathbf{\Upsilon }%
_{\mu \nu }=0$, we get to f\/irst order on f\/ields, with respect to N-adapted
bases (\ref{dder}) and (\ref{ddif}),
\begin{gather}
\widehat{\mathbf{R}}_{\alpha \beta } =0, \notag  \\
2 \widehat{\mathbf{D}}^{\nu }\mathbf{a}_{\mu \nu } =-\mathbf{W}_{\mu },
\notag \\
\big( \widehat{\square }+\mu ^{2}\big) \mathbf{a}_{\mu \nu } =2\widehat{%
\mathbf{R}}_{\cdot \ \nu \cdot \mu }^{\sigma \cdot \ \beta }\mathbf{a}%
_{\beta \sigma }+\frac{1}{3} \widehat{\mathbf{D}}_{[\mu }\mathbf{W}_{\nu ]},\label{vacnsmcg}
\end{gather}%
where $\mathbf{a}_{\alpha \beta }={}^{1}\mathbf{a}_{\alpha \beta }$, $\mathbf{W}%
_{\mu }={}^{1}\mathbf{W}_{\mu }$, $\widehat{\square }=\widehat{\mathbf{D}}^{\nu
}\widehat{\mathbf{D}}_{\nu }$ for $\widehat{\mathbf{D}}^{\nu }=\mathbf{g}%
^{\nu \mu }\widehat{\mathbf{D}}_{\mu },$ for $\mathbf{g}^{\nu \mu }$ being
inverse to $\mathbf{g}_{\alpha \beta }$ (these d-tensors are used for
rasing and lowering indices), and $\widehat{\mathbf{R}}_{\cdot \ \nu \cdot
\mu }^{\sigma \cdot \ \beta }$ and $\widehat{\mathbf{R}}_{\alpha \beta }$
are respectively curvature and Ricci d-tensors, with h- and
v-decompositions, def\/ined by formulas~(\ref{dcurvc}) and~(\ref{driccic}).
Following a d-tensor calculus similar to that in~\cite{legmoff,moff1a}, but for a canonical background d-connection~$\widehat{\mathbf{\Gamma }}
_{\ \beta \gamma }^{\alpha }$, we can represent (\ref{vacnsmcg}) in the form%
\begin{gather}
\widehat{\mathbf{R}}_{\alpha \beta } =0, \notag  \\
\ \widehat{\mathbf{D}}^{\sigma }\mathbf{a}_{\mu \sigma } =\frac{1}{\mu ^{2}%
}\ \widehat{\mathbf{D}}^{\nu }\big( 2\widehat{\mathbf{R}}_{\cdot \ \nu
\cdot \mu }^{\sigma \cdot \ \beta }\mathbf{a}_{\sigma \beta }-(\widehat{%
\mathbf{R}}\mathbf{a})_{\mu \nu }\big) ,  \notag \\
\big( \widehat{\square }+\mu ^{2}\big) \mathbf{a}_{\mu \nu } =\mathbf{M}%
_{\mu \nu },  \label{vacnsmcge}
\end{gather}%
where
\begin{equation*}
\mathbf{M}_{\mu \nu }=2\widehat{\mathbf{R}}_{\cdot \ \nu \cdot \mu }^{\sigma
\cdot \ \beta }\mathbf{a}_{\beta \sigma }+\frac{2}{\mu ^{2}}\ \widehat{%
\mathbf{D}}{}_{[\nu }\widehat{\mathbf{D}}^{\rho }\big( 2\widehat{\mathbf{R}}%
_{\cdot \ \rho \cdot \mu }^{\sigma \cdot \ \beta }\mathbf{a}_{\beta \sigma
}-(\widehat{\mathbf{R}}\mathbf{a})_{\mu ]\rho }\big)
\end{equation*}%
and $(\widehat{\mathbf{B}}\mathbf{a})_{\mu \nu }$ denotes additional terms
involving products of the Riemann d-tensor and skewsymmetric part of
d-metric.

There were constructed a number of solutions with $\widehat{\mathbf{R}}%
_{\alpha \beta }=0,$ see a review of results in~\cite{vrfg}, where it
is emphasized that we can constrain additionally the integral varieties of
these equations in order to generate of\/f-diagonal vacuum solutions in
general relativity. They can be used for gene\-ra\-li\-zations in NGT when the
background is nonholonomic and/or to model a~Finsler/Lag\-ran\-ge like
conf\/igurations. We constructed such spacetime models with running physical
constants and nonsymmetric metrics in~\cite{avnsm01}, where nontrivial
values for $\mathbf{a}_{\beta \sigma }$ where def\/ined from the nonholonomic
Ricci f\/low evolution equations. Certain classes of those solutions, when the
f\/low parameter is not identif\/ied with a time like coordinate can be
constrained additionally to def\/ine solutions of (\ref{vacnsmcge}). For
instance, the solitonic and pp-wave solutions with vanishing curvature $2
\widehat{\mathbf{R}}_{\cdot \ \nu \cdot \mu }^{\sigma \cdot \ \beta }$ at
asymptotics result positively in nontrivial solutions for $\mathbf{a}_{\mu
\sigma }$ which closely approximate solutions of Proca equations labelled
additionally by a Ricci f\/low parameter.

We conclude that we can perform nonholonomic deformation of symmetric
metrics into nonsymmetric ones following the method of anholonomic frames
and nonholonomic Ricci f\/lows. Such geometric conf\/igurations are also
admissible from the viewpoint of metric compatible/non\-com\-pa\-tib\-le NGT which
presents strong theoretical arguments for physical models with nonsymmetric
metrics and nonholonomic conf\/igurations.

\section[Gravity and nonsymmetric Lagrange-Finsler spaces]{Gravity and nonsymmetric Lagrange--Finsler spaces}
\label{snsfls}

The approach to geometrization of mechanics on tangent bundles
of the R.~Miron's school on generalized Lagrange, Hamilton and Finsler
geometry is strongly related to the geometry of nonlinear connections on
(co) vector/tangent bundles and their higher order generaliza\-tions~\cite{ma1987,ma,mhss} (this direction was developed as a generalization of the
geometry of Finsler and Cartan spaces). From formal point of view any
regular mechanics models can modelled as a Riemann--Cartan geometry with
nonholonomic distributions and, inversely, under well def\/ined conditions,
gravitational interactions admit an equivalent modelling by (semy) spary
conf\/igurations for an ef\/fective mechanics or nonlinear optics \cite{vrfg}.
Here, we note that our approach is dif\/ferent from the so-called analgous
gravity \cite{blv} where gravitational (for instance, black hole ef\/fects)
are ef\/fectively modelled by heuristic media and f\/lows, but not following a
rigorous geometric formalism, for instance, that of N-connections and
geometric mechanics.

The aim of this section is to prove that Lagrange--Finsler geometry and
nonholonomic gravity can be naturally related to NSG and def\/ine certain
models of nonsymmetric Lagrange--Finsler geometry.

Let us consider a regular Lagrangian $L(x,y)=L(x^{i},y^{a})$ modelled on $%
\mathbf{V}$, when the Lagrange metric (equivalently, Hessian){\samepage
\begin{equation}
{}^{L}g_{ij}=\frac{1}{2}\frac{\partial ^{2}L}{\partial y^{i}\partial y^{j}}
\label{lm}
\end{equation}%
is not degenerated, i.e.\ $\det |g_{ij}| \neq 0,$}

N-connections were f\/irst introduced in Finsler and Lagrange geometry by
considering (semi) spray conf\/igurations
\begin{equation}
\frac{dy^{a}}{d\varsigma }+2G^{a}(x,y)=0,  \label{ngeq}
\end{equation}
of a curve $x^{i}(\varsigma )$ with parameter $0\leq \varsigma \leq
\varsigma _{0}$, when $y^{i}=dx^{i}/d\varsigma $ [spray conf\/igurations are
obtained for integrable equations]. One holds the fundamental result (proof
is a straightforward computation):

\begin{theorem}
\label{t1}For
 $
4G^{j}={}^{L}g^{ij}\left( \frac{\partial ^{2}L}{\partial y^{i}\partial x^{k}}%
y^{k}-\frac{\partial L}{\partial x^{i}}\right)$,
with ${}^{L}g^{ij}$ inverse to ${}^{L}g_{ij},$ the ``nonlinear'' geodesic
equations \eqref{ngeq} are equivalent to the Euler--Lagrange equations
 $\frac{d}{d\varsigma }\left( \frac{\partial L}{\partial y^{i}}\right) -\frac{%
\partial L}{\partial x^{i}}=0$.
\end{theorem}

Finsler conf\/igurations can be obtained in a particular case when $%
L(x,y)=F^{2}(x,y)$ for a~homogeneous fundamental function $F(x,\lambda
y)=\lambda F(x,y)$, $\lambda \in \mathbb{R}.$ Lagrange and Finsler geometries
can be also modelled on N-anholonomic manifolds \cite{bejf,vrfg} provided,
for instance, with canonical N-connection structure
\begin{equation}
{}^{L}N_{i}^{a}=\frac{\partial G^{a}}{\partial y^{i}}.  \label{clnc}
\end{equation}

\begin{proposition}
A N-connection defines a set of nonholonomic preferred frames
\begin{equation*}
{}^{L}\mathbf{e}_{\alpha }=\left[ {}^{L}\mathbf{e}_{i}=\frac{\partial }{\partial x^{i}}-{}^{L}N_{i}^{a}(u)\frac{\partial }{\partial y^{a}},e_{b}=\frac{\partial }{\partial y^{b}}\right]
\end{equation*}
and coframes
\begin{equation*}
{}^{L}\mathbf{e}^{\alpha }=\big[e^{i}=dx^{i},\mathbf{e}^{a}=dy^{a}+{}^{L}N_{i}^{a}(x,y)dx^{i}\big].
\end{equation*}
\end{proposition}

\begin{proof}
One computes the nontrivial nonholonomy coef\/f\/icients ${}^{L}w_{ib}^{a}=\partial {}^{L}N_{i}^{a}/\partial y^{b}$ and ${}^{L}w_{ij}^{a}={}^{L}\Omega _{ji}^{a}={}^{L}\mathbf{e}_{i}{}^{L}N_{j}^{a}-{}^{L}\mathbf{e}_{j}{}^{L}N_{i}^{a}$ (where ${}^{L}\Omega _{ji}^{a}$ are the
coef\/f\/icients of the N-connection curvature) for
\begin{equation*}
\left[ {}^{L}\mathbf{e}_{\alpha },{}^{L}\mathbf{e}_{\beta }\right] ={}^{L}
\mathbf{e}_{\alpha }{}^{L}\mathbf{e}_{\beta }-{}^{L}\mathbf{e}_{\beta }{}^{L}
\mathbf{e}_{\alpha }={}^{L}w_{\alpha \beta }^{\gamma }{}^{L}\mathbf{e}
_{\gamma }.\tag*{\qed}
\end{equation*}\renewcommand{\qed}{}
\end{proof}

One holds:

\begin{theorem}
Any regular Lagrange mechanics $L(x,y)=L(x^{i},y^{a})$ can be modelled by
the geometry of a N-anholonomic manifold $\mathbf{V}^{n+n}$ enabled with
N-connection ${}^{L}\mathbf{N}$ and canonical metric structure%
\begin{equation}
{}^{L}\mathbf{g}={}^{L}g_{ij}(x,y)\left[ e^{i}\otimes e^{j}+{}^{L}\mathbf{e}%
^{i}\otimes {}^{L}\mathbf{e}^{j}\right] .  \label{m1l}
\end{equation}
\end{theorem}

\begin{proof}
For $\mathbf{V}=TM$, the metric (\ref{m1l}) is just the Sasaki lift of (\ref{lm}) on total space; see, for instance, \cite{ma}. In abstract form, such
canonical constructions can be performed similarly for any N-anholonomic
manifold $\mathbf{V.}$ This approach to geometric mechanics follows from the
fact that the (semi) spray conf\/igurations are related to the N-connection
structure and def\/ined both by the Lagrangian fundamental function and the
Euler--Lagrange equations, see Theorem~\ref{t1}.
\end{proof}

It is important:

\begin{proposition}
Any regular Lagrangian $L$ defines on $\mathbf{V}^{n+n}$ a preferred metric
compatible d-connection structures and metric compatible Lagrange (Finsler,
for $L=F^{2})$ canonical d-connection ${}^{L}\widehat{\mathbf{D}},$ or
normal d-connection ${}_{n}^{L}\mathbf{D},$ Cartan d-connection ${}_{c}^{L}%
\mathbf{D}$.
\end{proposition}

\begin{proof}
We can compute the coef\/f\/icients of these d-connections by introducing
formulas (\ref{m1l}) and (\ref{clnc}), respectively, in (\ref{candcon}), (\ref{ccd}) and (\ref{acscomp}).
\end{proof}

Because, in general, a regular Lagrange mechanics induces a nontrivial
N-connection structure, we can consider:

\begin{claim}
There are nonholonomic Ricci flows of Lagrange (Finsler) spaces resulting in
additional nonsymmetric components of d-metric~\eqref{m1l}.
\end{claim}

\begin{proof}
It follows from equation (29) in~\cite{avnsm01} $\frac{\partial }{\partial \chi }(N_{j}^{e}\underline{a}_{be})$, where $\chi $ is the Ricci
f\/low parameter. Ricci f\/lows of Lagrangians $L(\chi )$ induce f\/lows of~${}^{L}N_{j}^{e}(\chi ),$ which results in nontrivial values of~$\underline{a}%
_{be}$. The d-metric (\ref{m1l}) has to be extended to a variant of
d-metric (\ref{hvm}), when ${}^{L}\check{\mathbf{g}}={}^{L}\mathbf{g}+\mathbf{a.}$
\end{proof}

The above Claim is supported by

\begin{example}
A class of simplest examples of nonsymmetric Lagrange geometry is generated
by a regular Lagrangian $L(x,y)$ and any skew symmetric f\/ield $\mathbf{a}$
when ${}^{L}\check{\mathbf{g}}={}^{L}\mathbf{g}+\mathbf{a}$ and the
d-connection ${}_{\circ }\Gamma _{\ \beta \gamma }^{\alpha }=\big({}
_{\circ }L_{jk}^{i},{}_{\circ }C_{bc}^{a}\big) $ from Example~\ref{ex1} is
taken to be the Cartan d-connection, i.e.~${}_{\circ }\Gamma _{\ \beta \gamma
}^{\alpha }={}_{c}^{L}\Gamma _{\ \beta \gamma }^{\alpha }$ and the
general metric compatible d-connections on $\mathbf{V}^{n+n}$ are def\/ined
in the form ${}_{\ast }^{L}\Gamma _{\ \beta \gamma }^{\alpha }=\big( {}
_{\ast }^{L}L_{jk}^{i},{}_{\ast }^{L}C_{bc}^{a}\big) $ where
\begin{gather}
{}_{\ast }^{L}L_{jk}^{i} ={} _{c}^{L}L_{jk}^{i}+\frac{1}{2}\big[{}^{L}g^{ir}{}_{c}^{L}D_{k}{}^{L}g_{rj}+
{}_{L}^{\pm }O_{sj}^{ir}\big(\check{a}^{st}{}_{c}^{L}D_{k}a_{tr}+3{l}%
_{t}^{s}{}_{c}^{L}D_{k}{l}_{r}^{t}-{}_{c}^{L}D_{k}{l}_{r}^{s}\big)\big], \notag  \\
{}_{\ast }^{L}C_{bc}^{a} ={}_{c}^{L}C_{bc}^{a}+\frac{1}{2}\big[{}^{L}g^{ah}{}_{c}^{L}D_{c}{}^{L}g_{hb}+
{}_{L}^{\pm }O_{eb}^{ah}\big(\check{a}^{ed}{}_{c}^{L}D_{c}a_{dh}+3{l}_{d}^{e}{}_{c}^{L}D_{c}{l}_{h}^{d}-{}_{c}^{L}D_{c}{l}%
_{h}^{e}\big)\big]  \label{aux11a}
\end{gather}%
is d-metric compatible, i.e.\ satisf\/ies the conditions ${}_{\ast }^{L}\mathbf{D}^{\mathbf{L}}\check{\mathbf{g}}=0$; we put a left label ``L'' on values generated by~$L$.
We note that similar constructions are possible for any metric compatible
d-connection generated by $L$, including the class of normal ones, or the
canonical d-connection.
\end{example}

For more general constructions, we formulate:

\begin{theorem}
The set of Lagrange (Finsler) spaces with nonsymmetric metric compatible
d-connections ${}^{L}\mathbf{\Gamma }_{\ \beta \gamma }^{\alpha }
=\big( {}^{L}L_{jk}^{i},{}^{L}C_{bc}^{a}\big) $ generated by a regular
Lagrangian $L(x,y)$ (fundamental Finsler funciton $F(x,y),$ where $L=F^{2})$
on  $\mathbf{V}^{n+n}$ is parametrized by N-adapted coefficients
\begin{equation}
{}^{L}L_{jk}^{i}={}_{\ast }^{L}L_{jk}^{i}+{}_{L}^{+}O_{sj}^{ir}{}_{L}^{+}\Psi _{pr}^{sm}  \mathbf{Y}_{mk}^{p}
\qquad \mbox{and}\qquad {}^{L}C_{bc}^{a}={}_{\ast }^{L}C_{bc}^{a}+{}_{L}^{+}O_{eb}^{ah}{}_{L}^{+}\Psi _{fh}^{ed}
\mathbf{Z}_{dc}^{f},  \label{auxa}
\end{equation}%
where ${}_{\ast }^{L}L_{jk}^{i}$ and ${}_{\ast }^{L}C_{bc}^{a}$ are given
respectively by formulas \eqref{aux11a}; $_{hL}^{+}O=\{{}_{L}^{+}O_{sj}^{ir}\}$ and ${}_{hL}^{+}O=\{{}_{L}^{+}O_{eb}^{ah}\}$ are
defined by formulas \eqref{obop} and \eqref{aux4} but for ${}^{L}\mathbf{g}$
\eqref{m1l}; ${}_{hL}^{+}\Psi =\{{}_{L}^{+}\Psi _{pr}^{sm}\}$ and ${}_{vL}^{+}\Psi =\{{}_{L}^{+}\Psi _{fh}^{ed}\}$ are defined by the
skewsymmetric part of metric $\mathbf{a}$ following formulas \eqref{aux5}
and \eqref{aux6} assuming that the d-metric ${}^{L}\check{\mathbf{g}}={}^{L}%
\mathbf{g}+\mathbf{a}$ is natural, i.e.\ the symmetric and skewsymmetric Obata
operators satisfy the conditions \eqref{obcr}, and $\mathbf{Y}=
\{Y_{mk}^{p}\} $ and $\mathbf{Z}=\{Z_{dc}^{f}\}$ are correspondingly
arbitrary horizontal and vertical d-tensors of type $(1,2)$.
\end{theorem}

\begin{proof}
The statements of this theorem follow from Corollary \ref{corel}, Theorem~\ref{thmain} and Conclusion~\ref{conclthmain} specif\/ied for d-metrics ${}^{L}\check{\mathbf{g}}={}^{L}\mathbf{g}+\mathbf{a}$, and metric compatible
d-connections (\ref{aux11a}) and~(\ref{auxa}).
\end{proof}

The last main result, in this work, is that we can model various classes of
generalized Lagrange--Finsler spaces in NGT and, inversely, we can model as
ef\/fective regular Lagrange systems (or Finsler conf\/igurations) the f\/ield
interactions in NGT:

\begin{result}
A generalized Lagrange (Finsler) geometry defined by a natural ${}^{L}%
\check{\mathbf{g}}={}^{L}\mathbf{g}+\mathbf{a}$, with~${}^{L}\mathbf{g}$  defined
by \eqref{m1l}, canonical N-connection ${}^{L}\mathbf{N}$ \eqref{clnc} and
a metric compatible d-connection is nonholonomically equivalent to a NGT
model if the mentioned geometric d-tensor and connections satisfy the
``nonsymmetric'' field equations~\eqref{nscgfe}.
\end{result}

\begin{proof}
The direct statement is obvious. The inverse one can be supported by some
examples of solutions. The length of this article does not allow us to
present details on constructing such solutions; see examples, detailed
discussions and references in Sections~5 and~A.5 of~\cite{vrfg} and~\cite{vncg} and certain solutions for f\/ixed Ricci f\/low parameter in \cite{vrfsol1,vvisrf1,vvisrf2,vrf04,vrf05,avnsm01}. For any class of those
solutions modelling solionic pp-waves, black ellipsoids/tori, locally
anisotropic Taub NUT spaces \dots having the property that $\widehat{\mathbf{R}
}_{\alpha \beta }=0$ (def\/ining vacuum background solutions for the canonical
d-connection) and $\widehat{\mathbf{R}}_{\cdot  \nu \cdot \mu }^{\sigma
\cdot  \beta }\rightarrow 0$, let say for a radial/cilinder coordinate \mbox{$r\rightarrow \infty $}, in~(\ref{vacnsmcge}). For such conf\/igurations, we can
approximate the vacuum NGT f\/ield equations for the skewsymmetric part of
metric as $\big( \widehat{\square }+\mu ^{2}\big) \mathbf{a}_{\mu \nu }=0$
and $\widehat{\mathbf{D}}^{\sigma }\mathbf{a}_{\mu \sigma }=0$ which are
well the known Proca f\/ield equations. Technically, it is a very dif\/f\/icult
task to construct exact solutions in NGT, but it is possible almost always
to construct certain approximations proving the existence of such solutions.
This states positively that certain classes of symmetric and nonsymmetric
Lagrange--Finsler conf\/igurations can be extracted from NSG and, inversely,
such regular mechanical systems can be used for modelling (non)symmetric
gravitational interac\-tions.\end{proof}

Finally, we note that similar Results can be formulated, for instance, for
noncommutative and quantum geometric generalizations of Lagrange--Finsler
geometry and NGT which are topics of our further investigations.

\section{Conclusions and discussion}\label{sec8}

In summary, we elaborated a geometric approach to physical theories on
spacetimes provided with nonlinear and linear connections compatible with
nonsymmetric metrics, in the context of the geometry of nonholonomic
Riem\-ann--Cartan manifolds and generalized Lagrange--Finsler spaces. Toward
such results, we have applied a programme of research that is based prior on
the moving anholonomic frame method and nonholonomic deformations of
geometric structures, nonlinear connection (N-connection) formalism,
metri\-zation procedure of distinguished connections (d-connections), i.e.\
the linear connections adapted to a N-connections structure, and former
results from the nonsymmetric gravity theory (NGT). The premise of this
methodology is that one can be constructed certain classes of geometric
models of NGT which are metric compatible and satisfy all conditions imposed
for the modern paradigm of standard physics (such theories are  well
def\/ined in the linear approximation on Minkowski space and  expanded
about arbitrary Einstein, or about/to Lagrange--Finsler and other type  backgrounds).
The validity of this approach was substantiated by reproducing and
understanding a number of ef\/fects in modern gravity and cosmology.

Nonholonomic distributions and nonsymmetric metrics arise naturally in: the
theory of nonholonomic Ricci f\/lows, as solutions of the evolution equations;
 dynamics of  constrained physical systems; geometry of
nonholonomic maps and deformations of geometric structures on classical and
geometrically quantized models of gravity; noncommutative and quantum group
deformations of gravitational, gauge and spinor theories. Formulation of a
rigorous geometric approach to the theory of classical and quantum f\/ield
interactions and f\/low evolution equations with generic of\/f-diagonal metrics
and constraints positively requests a detailed study of `nonsymmetric and
nonholonomic theories'.

Historically, the f\/irsts theoretical schemes for theories
with nonsymmetric metrics and conceptual and heuristic arguments to suit
with existing experimental data were proposed in~\cite{nseinst1,nseinst,nseisenh1,nseisenh2,moffrev,nsgtjmp}. That stimulated
certain interest of the scientif\/ic community (together with more or less motivated criticism) and a lot of
discussions how to cure such theories \cite{ddmcc,prok,vstabns}.  More
recently, new evidences coming from the Ricci f\/low theory, deformation
quantization, geometric methods of constructing exact solutions of
gravitational f\/ield and evolution equations, string gravity and nonholonomic
spinor and Clif\/ford--Lie algebroids etc has revealed a quite unexpected and additional
theoretical support for NGT (for simplicity, in this work, we do not
concern various speculations on some cosmological and astrophysical
evidences, see details in~\cite{moff0505326,prok,mofft}).

Let us provide and discuss  contemporary motivations  for
theories with nonsymmetric metrics and nonholonomic structures elaborated in
a metric compatible form:
\begin{enumerate}\itemsep=0pt
\item Nonsymmetric metrics are positively induced by Ricci f\/low
evolutions of (pseudo) Riemannian/Einstein metrics. Such results were proven
following  methods of nonholonomic geometric analysis and by a number of
examples  of exact
solutions def\/ining Ricci f\/lows of valuable physical equations \cite%
{vrf02,avnsm01}. The condition of metric compatibility was crucial for such
constructions: it is not clear how to generalize the Perelman's functionals
(with corresponding entropy, average energy and analogous thermodynamical
values which can be def\/ined even for gravitational and mechanical motion/f\/ield
equations) and relevant conservation laws for spaces with nonmetricity. If
 we try to elaborate certain nonmetric geometric and physical (?)
generalizations of former metric compatible constructions, this is possible only as distorsions from some well def\/ined
metric compatible conf\/igurations. So, metric compatibility is a crucial
condition even we are oriented to develop nonmetric geometric and physical
theories. Here we note that nonsymmetric metrics and metric compatible connections, in
general, on nonholonomic manifolds, and their dynamics and evolution under
geometric f\/lows are not prohibited by any general theoretical models or experimental data
in modern physics.

\item We proved that the Einstein gravity can be re-written equivalently  in
so--called almost K\"{a}hler variables, and related Lagrange--Finsler
variables, which is used for a study of stability of possible nonsymmetric
metric generalizations \cite{vstabns}. But such stable conf\/igurations seem
to exist only for some classes of nonholonomic constraints and under the condition of metric
compatibility if we wont to preserve a physical limit to general relativity.

\item Nonsymmetric metric conf\/igurations are derived also in noncommutative
geometry and gravity and  string/brane gravity \cite{moffncqg,vggr,vqgr3,castro1}. Following dif\/ferent geometric techniques,
introducing  anti-commuting/noncommuting variables, performing Seiberg--Witten transforms
in gau\-ge gravity, or using Born's reciprocity principle etc, noncommutative
and nonsymmetric metrics and connections are generated to be mutually
related as certain f\/ield variables. In nonholonomic and metric compatible
form such conf\/igurations can be def\/ined as exact solutions in Einstein,
string and gauge gravity theories with noncommutative variables.

\item Applying methods of Fedosov quantization to Einstein, gauge and string
gravity theories, as well to Lagrange--Finsler--Hamilton systems, deforming  an ef\/fective symplectic (gravitational) form, and related symplectic
connections, we may def\/ine quantum corrections as nonsymmetric metric
contributions, see details in \cite{vqgr3,vdqlfs,castro} and references
therein. Such quantum systems are (in general)  nonholonomic and
characterized by certain canonical symmetries and invariants if the
constructions are metric compatible.

\item The most striking revelation of nonholonomic nonsymmetric (commutative
and/or noncommutative) theories is that a corresponding generalization of the
Dirac operator/equa\-tions and inherent nonhlonomic spinor/Clif\/ford structures
have to be elaborated. Such constructions were performed in a series of our
previous works \cite{vfs,vhs,vclalg}, for spinors in Lagrange--Finsler (and
higher order generalizations) and Clif\/ford--Lie algebroids, with further
developments in nonholonomic metric-af\/f\/ine and noncommutative geometry (see
Parts I and III in~\cite{vsgg}) and for spectral functionals and
noncommutative Ricci f\/lows \cite{vncdoconnes}. In all cases, the condition
of metric compatibility is a cornstone one for physical viable and simplest
theoretical/geometric generalizations of classical and quantum f\/ield theories. For instance, it is a problem to def\/ine spinors even locally and propose a well def\/ined generalization of Dirac equations for metric noncompatible manifolds.

\item \looseness=1 Incorporating the above mentioned results into standard theories of
physics (the problem is~also discussed in details in~\cite{vrfg} and
Introduction section to \cite{vsgg}), we have to solve a~ge\-ne\-ral
mathematical problem how to model geometrically various classes of spaces enabled with
nonsymmetric metrics and  nonholonomic distributions. How to def\/ine
conservation laws and relevant invariants which would characterize such
theories (in NGT, the problem was discussed, for instance, in~\cite{legmoff} and for the case of nonholonomic spaces in~\cite{vsgg,vrfg,vrf02})? It should be emphasized that such geometric structures
exist even in Einstein gravity if solutions with generic of\/f-diagonal
metrics and nonholonomic frames are introduced into consideration. So, a standard approach both to generalized f\/ield equations and nonholomic constraints and well def\/ined procedures of quantization of nonlinear f\/ield theories and corresponding conservation laws are possible only if the condition of metricity is satisf\/ied.
\end{enumerate}

Following the arguments and results outlined above in points 1-6, we
conclude that the presented in this work study of the geometry of
nonholonomic manifolds endowed with compatible nonsymmetric metric and
linear and nonlinear connection structures is  motivated not only by
certain ``academic'' generalizations of metric and connection structures in dif\/ferential geometry but also by a series of results and requests from modern theoretical and mathematical physics.

Let us speculate on some further perspectives and applications of geometric methods from the theory of nonholonomic manifolds provided with compatible nonsymmetric metric and nonlinear and linear connections structures:

In the approximation of week skewsymmetric part of metrics, we can generate
various gravity models with ef\/fective scalar/vector/tensor interactions
which seem to propose original solutions of the problem of dark matter and
connections to nonhomogeneous and locally anisotropic cosmological models
and suggest new tests of gravity. It is possible to construct in explicit form
various classes of exact and approximate solutions with variation/runing of
physical constants and their nonlinear and (locally) anisotropic
polarizations. Working with nonholonomic distributions and associated
N-connections, one can be extracted from NGT dif\/ferent models of Lagrange
and Finsler geometries and their generalizations. We also argue that we can
model by regular mechanical systems certain classes of (non)symmetric and/or
(non)holonomic gravitational interactions and, inversely, the geometric
mechanics can be represented and generalized as certain classes of
nonsymmetric and/or nonholonomic Riemann--Cartan spaces.

In this work, we have shown how NGT models can be elaborated in the form
when there are satisf\/ied the metric compatibility and N-connection adapted
conditions. This is very important both from conceptual and technical points
of views. For instance, working with metric compatible connections, we can
develop the theory of Clif\/ford structures with nonsymmetric metrics and
N-connections and formulate the criteria when global spinor constructions
for such spaces are possible. We are able to compute the topological
obstructions for constructing such nonsymmetric spinor generalizations and
suggested variants of extending such theories. It is also possible to def\/ine
self-consistently the theory of Dirac operators with further extensions to
noncommutative models and deformation quantization of NGT. All such tasks
become less physical and realistic, and very dif\/f\/icult to be solved
mathematically, if we work with metric noncompatible connections.

Finally, it should be emphasized that using canonical metric compatible connections,
even (in general) the metrics can be nonsymmetric, we can generalize the
anholonomic frame method of constructing exact and approximate solutions in
various models of gravity. The obtained in this works results and mentioned
theoretical and phenomenological directions provide a considerable setback
for our forthcoming research projects.


\appendix

\section{Some important local formulas}\label{app}

The N-adapted splitting into h- and v-covariant derivatives is stated by
\begin{equation*}
h\mathbf{D}=\{\mathbf{D}_{k}=\left( L_{jk}^{i},L_{bk\;}^{a}\right) \} \qquad
\mbox{and}\qquad v\mathbf{D}=\{\mathbf{D}_{c}=\left( C_{jc}^{i},C_{bc}^{a}\right) \},
\end{equation*}%
where, by def\/inition,
 $L_{jk}^{i}=\left( \mathbf{D}_{k}\mathbf{e}_{j}\right) \rfloor e^{i}$, $L_{bk}^{a}=\left( \mathbf{D}_{k}e_{b}\right) \rfloor \mathbf{e}%
^{a}$, $C_{jc}^{i}=\left( \mathbf{D}_{c}\mathbf{e}_{j}\right) \rfloor
e^{i}$, $C_{bc}^{a}=\left( \mathbf{D}_{c}e_{b}\right) \rfloor \mathbf{e}%
^{a} $.
The components $\mathbf{\Gamma }_{\ \alpha \beta }^{\gamma }=\big(
L_{jk}^{i},L_{bk}^{a},C_{jc}^{i},C_{bc}^{a}\big) $ completely def\/ine a
d-connection $\mathbf{D}$ on a N-anholonomic manifold $\mathbf{V}$.

The simplest way to perform computations with d-connections is to use
N-adapted dif\/ferential forms like
$\mathbf{\Gamma }_{\ \beta }^{\alpha }=\mathbf{\Gamma }_{\ \beta \gamma
}^{\alpha }\mathbf{e}^{\gamma }$,
 with the coef\/f\/icients def\/ined with respect to (\ref{ddif}) and (\ref{dder}).
For instance, the N-adapted coef\/f\/icients of torsion (\ref{tors}), i.e.\
d-torsion, is computed in the form
 $
\mathcal{T}^{\alpha }\doteqdot \mathbf{De}^{\alpha }=d\mathbf{e}^{\alpha
}+\Gamma _{\ \beta }^{\alpha }\wedge \mathbf{e}^{\beta },
 $
where
\begin{gather}
T_{\ jk}^{i} = L_{\ jk}^{i}-L_{\ kj}^{i},\qquad T_{\ ja}^{i}=C_{\ ja}^{i},\qquad T_{\
ji}^{a}=\Omega _{\ ji}^{a},\nonumber\\
T_{\ bi}^{a} =\frac{\partial N_{i}^{a}}{\partial y^{b}}-L_{\ bi}^{a},\qquad T_{\ bc}^{a}=C_{\ bc}^{a}-C_{\ cb}^{a},  \label{dtorsc}
\end{gather}%
 for $\Omega _{\ ji}^{a}$ being the curvature of N-connection (\ref{ncurv}).

By a straightforward d-form calculus, we can f\/ind the N-adapted components
of the curvature~(\ref{curv}) of a d-connection $\mathbf{D}$,
 $
\mathcal{R}_{~\beta }^{\alpha }\doteqdot \mathbf{D\Gamma }_{\ \beta
}^{\alpha }=d\mathbf{\Gamma }_{\ \beta }^{\alpha }-\mathbf{\Gamma }_{\ \beta
}^{\gamma }\wedge \mathbf{\Gamma }_{\ \gamma }^{\alpha }=\mathbf{R}_{\ \beta
\gamma \delta }^{\alpha }\mathbf{e}^{\gamma }\wedge \mathbf{e}^{\delta },$
 i.e. the d-curvature,
\begin{gather}
R_{\ hjk}^{i} =\mathbf{e}_{k}\left( L_{\ hj}^{i}\right) -\mathbf{e}%
_{j}\left( L_{\ hk}^{i}\right) +L_{\ hj}^{m}L_{\ mk}^{i}-L_{\ hk}^{m}L_{\
mj}^{i}-C_{\ ha}^{i}\Omega _{\ kj}^{a},  \notag \\
R_{\ bjk}^{a} =\mathbf{e}_{k}\left( L_{\ bj}^{a}\right) -\mathbf{e}%
_{j}\left( L_{\ bk}^{a}\right) +L_{\ bj}^{c}L_{\ ck}^{a}-L_{\ bk}^{c}L_{\
cj}^{a}-C_{\ bc}^{a}\Omega _{\ kj}^{c},   \notag \\
R_{\ jka}^{i} =e_{a}L_{\ jk}^{i}-\mathbf{D}_{k}C_{\ ja}^{i}+C_{\
jb}^{i}T_{\ ka}^{b},\qquad  R_{\ bka}^{c} = e_{a}L_{\ bk}^{c}-\mathbf{D}_{k}C_{\ ba}^{c}+C_{\
bd}^{c}T_{\ ka}^{c},  \notag \\
R_{\ jbc}^{i} =e_{c}C_{\ jb}^{i}-e_{b}C_{\ jc}^{i}+C_{\ jb}^{h}C_{\
hc}^{i}-C_{\ jc}^{h}C_{\ hb}^{i},\notag\\
R_{\ bcd}^{a} = e_{d}C_{\ bc}^{a}-e_{c}C_{\ bd}^{a}+C_{\ bc}^{e}C_{\
ed}^{a}-C_{\ bd}^{e}C_{\ ec}^{a}. \label{dcurvc}
\end{gather}

Contracting respectively the components of (\ref{dcurvc}), one proves that
the Ricci tensor $\mathbf{R}_{\alpha \beta }\doteqdot \mathbf{R}_{\ \alpha
\beta \tau }^{\tau }$ is characterized by h- v-components, i.e. d-tensors,%
\begin{equation}
R_{ij}\doteqdot R_{\ ijk}^{k},\qquad R_{ia}\doteqdot -R_{\ ika}^{k},\qquad
R_{ai}\doteqdot R_{\ aib}^{b},\qquad R_{ab}\doteqdot R_{\ abc}^{c}.
\label{driccic}
\end{equation}

Finally, we note that the def\/inition of scalar curvature requests a metric
structure, which is an additional geometric structure with respect to that
of d-connection.

\subsection*{Acknowledgements}

The work is performed during a visit at Fields
Institute. Author is grateful to Professors M.~Anastasiei and J.~Mof\/fat for kind support.

\pdfbookmark[1]{References}{ref}
\LastPageEnding

\end{document}